\documentclass[nanomaterials,article,submit,moreauthors,pdftex]{mdpi} 
\preto{\abstractkeywords}{\nolinenumbers}

\firstpage{1} 
\pubvolume{1}
\issuenum{1}
\articlenumber{0}
\pubyear{2021}
\copyrightyear{2020}
\datereceived{} 
\dateaccepted{} 
\datepublished{} 
\hreflink{https://doi.org/} 

\usepackage{physics}

\let\oldAA\AA
\renewcommand{\AA}{\text{\normalfont\oldAA}}
\Title{The Cost of Improving the Precision 
of the Variational Quantum Eigensolver 
for Quantum Chemistry}

\TitleCitation{The Cost of Improving the Precision 
of the Variational Quantum Eigensolver 
for Quantum Chemistry}

\Author{Ivana Miháliková $^{1,2,3
}$\orcidA{}, 
Matej Pivoluska $^{2,4}$*\orcidC{},
Martin Plesch $^{2,4}$\orcidB{},
Martin Friák $^{1,3,5}$\orcidD{},
Daniel Nagaj $^{2}$\orcidF{}, and
Mojmír Šob $^{5,1}$\orcidG{}
}

\AuthorNames{Ivana Miháliková, Martin Plesch, Matej Pivoluska, Martin Friák, Martin Saip, Daniel Nagaj, and Mojmír Šob}

\AuthorCitation{Miháliková, Ivana; et al.}

\address{%
$^{1}$ \quad Institute of Physics of Materials, Czech Academy of Sciences, Žižkova 22, CZ-616 62 Brno, Czech Republic \\
$^{2}$ \quad Institute of Computer Science, Masaryk University, Šumavská 416, CZ-602 00 Brno, Czech Republic\\
$^{3}$ \quad Department of Condensed Matter Physics, Faculty of Science, Masaryk University, Kotlářská 2, CZ-611 37 Brno, Czech Republic\\
$^{4}$ \quad Institute of Physics, Slovak Academy of Sciences, Dúbravská cesta 9, SK-841 04 Bratislava, Slovakia\\
$^{5}$ \quad Department of Chemistry, Masaryk University, Kotlářská 2, CZ-611 37 Brno, Czech Republic}

\corres{Correspondence: fyzimapi@savba.sk}

\abstract{
Quantum computing brings a promise of new approaches into computational quantum chemistry. While universal, fault-tolerant quantum computers are still not available, we want to utilize today's noisy quantum processors. One of their flagship applications is the variational quantum eigensolver (VQE) -- an algorithm to calculate the minimum energy of a physical Hamiltonian. In this study, we investigate how various types of errors affect the VQE, and how to efficiently use the available resources to produce precise computational results.
We utilize a simulator of a noisy quantum device, an exact statevector simulator, as well as physical quantum hardware to study the VQE algorithm for molecular hydrogen. We find that the optimal way of running the hybrid classical-quantum optimization is to (i) allow some noise in intermediate energy evaluations, using fewer shots per step and fewer optimization iterations, but require high final readout precision, (ii) emphasize efficient problem encoding and ansatz parametrization, and (iii) run all experiments within a short time-frame, avoiding parameter drift with time.
Nevertheless, current publicly available quantum resources are still very noisy and scarce/expensive, and even when using them efficiently it is quite difficult to obtain trustworthy calculations of molecular energies.
}


\begin{document}


\section{Introduction}

Computer simulations of quantum systems constitute a crucial tool for a deeper understanding of behaviour and properties of matter on the atomic scale. However, when investigating the electronic structure of larger molecules, we quickly encounter the limits of classical computers. The space and time requirements for  describing the states, and even more so for performing optimization on them, grow prohibitively. Thus, we resort to many clever types of approximations \cite{doi:10.1063/1.5129672, ONMethods, Qmontecarlo,Friesner-2005,Trygve-2008,Cremer-2011,Lyakh-2012,Yu-2016,Narbe-2017}. Oftentimes, though, they are not good enough, or start to scale badly.

It is natural to imagine that using one quantum mechanical system to simulate another could be a more efficient approach. The concept of a quantum computer, based on laws of quantum physics, was first proposed by Richard Feynman \cite{Feynman}. He envisioned that the way to deal with the exponential amount of information appropriate to study physical systems was to use quantum systems as computers themselves. We have come a long way since then, and today we have access to the first small, programmable quantum chips, together with 25 years of development of quantum algorithms \cite{algZoo,2018arXiv180403719A,10.5555/1972505} and protocols for simulation, optimization, and many other applications in sensing, cryptography and communication. \cite{RevModPhys.89.035002,RevModPhys.74.145,RevModPhys.92.025002,RevModPhys.89.015004}.

While fault-tolerant, universal quantum computing is still out of reach, much of today's development focuses at demonstrating quantum supremacy on problems without direct applications \cite{Arute2019,PhysRevLett.127.180501,PhysRevLett.127.180502}. Our goal here is less ambitious, but more practical, utilizing the imperfect computers we have at hand for a practical task. We investigate one of the key promised applications of quantum computing: the understanding of molecular structure. While the long-term goal is to find more efficient solutions for a problem deemed intractable on conventional computers for large problem sizes, we want to understand how well we can do today for small test cases.

The basic task is to solve the many-body Schrödinger equation

\begin{equation}\label{SCH}
\hat{H} \ket{\psi}  = \textit{E} \, \ket{\psi},
\end{equation}

\noindent
where $\hat{H}$ denotes the Hamiltonian operator, $\ket{\psi}$ is the system wave function and $\textit{E}$ represents the energy of the system. Finding the ground state energy of the studied system is a difficult optimization problem. However, we now know several promising approaches for applications in quantum chemistry \cite{Cao2019}, relying on natural quantum encodings of the problems, and algorithms utilizing superpositions and entanglement, resulting in efficient search and energy evaluation. 

One way of approximately obtaining the minimum eigenvalue of a Hamiltonian is to start with an initial guess and iteratively search for optimal parameters. One such popular approach aimed at solving the many-particle problem is the hybrid classical-quantum {\em Variational Quantum Eigensolver} (VQE) algorithm proposed by Peruzzo et al. \cite{vqe}. It is an application of the time-independent variational principle that benefits from cooperation of classical and quantum computers, and is suitable for use on near-term, imperfect quantum devices. For example, in \cite{vqe}, the authors used a small photonic quantum computer to calculate the He–H$^+$ system's ground state energies for various atomic distances.
Next, O'Malley et al.~\cite{OMalley-PRX-2016} studied the properties of the H$_2$ molecule using Google’s digital quantum computer with superconducting qubits. The researchers used two different methods for finding the ground-state energies of H$_2$: the phase-estimation algorithm (PEA), and the variational quantum eigensolver (VQE). The former method can in principle get the answer with arbitrary precision, but only if there are no errors in the process. As in practice, errors are always present, the VQE method works better. However, it also has its pitfalls, which we aim to elucidate.

In this work, we want to understand the possible sources of errors in VQE calculations when implemented on current, publicly available superconducting quantum processors. Our goal is to provide a recipe for efficient use of the scarce resources, as well as practical guidance for the starting practitioner, wanting to run quantum chemistry computations on current devices.



If our (superconducting) qubits were a closed system, the dynamical evolution of their state would be fully determined by their initial state and Hamiltonian. In reality, the system is partially open, the qubits interact with their environment, their internal interactions are not perfect, and so the overall state is not deterministic. A generalized theory of the interactions between a quantum state and its environment was derived by Bloch~\cite{Bloch-1957}, manifesting as extra degrees of freedom we cannot control. These affect the state of the system and result in a loss of coherence \cite{Krantz-APR-2019}.
Noise comes through controlling the qubits (e.g. via pulses applied to the qubits) or reading out the final states of the qubits (there is a finite probability that the recorded classical bit value will be flipped from the true outcome of a measurement). We also get errors from random fluctuations of parameters that are coupled to our qubits, such as (i) thermal voltage and current fluctuations in control lines or (ii) randomly fluctuating electric and magnetic fields in the local qubit environment. 

Minimizing various sources of noise and errors is a very complex and demanding task involving materials science, fabrication engineering, electronics design, cryogenic engineering, and qubit design.
Of course, our hope is fault-tolerant quantum processing, which is currently out of reach, thanks to precision and overhead requirements \cite{Preskill2018quantumcomputingin}. 
Meanwhile, we often employ methods for dealing with noisy data.
There are many different approaches to correcting the noisy results (see Refs. \cite{kandala,Kandala2019,PhysRevLett.119.180509,PhysRevX.8.031027, Geller_2021,mitigation2,Nachman-npjQI-2020,Maciejewski2020mitigationofreadout,Cai2021quantumerror,Suchsland2021algorithmicerror} and references therein), based on different techniques of noise characterization.  

Finally, even hypothetical noiseless quantum processors are prone to {\em stochastic noise}. Our calculations end with estimating the values of Pauli observables in the final state. This can only be done via sampling the outcome probabilities, with a limited number of shots. 
Moreover, stochastic noise comes in also through the heuristic nature of the classical optimizer part of the VQE algorithm. The good news is that we can investigate and control this noise, thanks to our selection of optimizer, and the parameters of the experimental runs. We will now attempt to characterize in detail how each source of noise influences the convergence of the VQE algorithm. We will then look at how to efficiently distribute our resources in order to achieve the desired precision of the calculations, if it is possible at all.

\section{Methods}

There are several preparatory steps before studying molecules on real quantum devices. We start with the molecular Hamiltonian in the second quantized form \cite{SQ}. The fermionic Hamiltonian with one and two-electron terms ($h_{ij}$ and $h_{ijkl}$) is given by

\begin{equation}\label{SQ}
	\hat{H} = \sum_{ij} h_{ij} \,\, a_{i}^\dagger a_{j} + \frac{1}{2} \sum_{ijkl} h_{ijkl} \,\, a_{i}^\dagger a_{j}^\dagger a_{l} a_{k},
\end{equation}

\noindent
where $a_{i}^\dagger$ and $a_{j}$ are fermionic creation and annihilation operators. We then need to map this fermionic Hamiltonian into qubit operators represented in the Pauli operator basis \cite{map}. This can be done in multiple ways, e.g. with the Bravyi-Kitaev \cite{bravyi} or Jordan-Wigner \cite{Jordan1928} transformations.
Here we choose the Bravyi-Kitaev transformation \cite{bravyi} of the hydrogen molecule Hamiltonian, produced using the publicly available {\em qiskit} package \cite{qiskit}. 
The code for this, and all of our following calculations is located at \cite{imihalik}. We also invite the reader to read the introductory VQE tutorial \cite{VQETut}. 

Choosing the {\em STO-3G} basis set, and the distance between hydrogen atoms set to $0.725 \; \AA$, not taking into account the Coulomb repulsiom between nuclei,
we arrive at a~$4$-qubit Hamiltonian 

\begin{align}
	\hat{H}_{\mathrm{H_2}}  &=  c_{\mathrm 0} \; \boldsymbol{1} 
	+ c_{\mathrm 1} \; Z_{\mathrm 0} 
	+ c_{\mathrm 2} \;  Z_{\mathrm 1}Z_{\mathrm 0}  
	+ c_{\mathrm 1}  \;  Z_{\mathrm 2}
	+ c_{\mathrm 2}  \;  Z_{\mathrm 3}Z_{\mathrm 2}Z_{\mathrm 1}
	+ c_{\mathrm 3}  \;  Z_{\mathrm 1}
	+ c_{\mathrm 4}  \;  Z_{\mathrm 2}Z_{\mathrm 0} \nonumber\\
	&+ c_{\mathrm 5}  \;  X_{\mathrm 2}Z_{\mathrm 1}X_{\mathrm 0}
	+ c_{\mathrm 6}  \;  Z_{\mathrm 3}X_{\mathrm 2}X_{\mathrm 0}
	+ c_{\mathrm 6}  \;  X_{\mathrm 2}X_{\mathrm 0} 
	+ c_{\mathrm 5}  \;  Z_{\mathrm 3}X_{\mathrm 2}Z_{\mathrm 1}X_{\mathrm 0} \label{H2_4_qubits}\\
	&+ c_{\mathrm 7}  \;  Z_{\mathrm 3}Z_{\mathrm 2}Z_{\mathrm 1}Z_{\mathrm 0}
	+ c_{\mathrm 7}  \;  Z_{\mathrm 2}Z_{\mathrm 1}Z_{\mathrm 0}
	+ c_{\mathrm 8}  \;  Z_{\mathrm 3}Z_{\mathrm 2}Z_{\mathrm 0}
	+ c_{\mathrm 3}  \;  Z_{\mathrm 3}Z_{\mathrm 1}. \nonumber 
\end{align}

\noindent
The $Z$ and $X$ terms are Pauli operators, and the coefficients $c_i$ are integrals calculated using the {\em qiskit.chemistry} package \cite{qiskit.chemistry}:

\begin{equation}
    \begin{array}{lll}
    c_{\rm 0} = -0.80718, \hskip1cm & c_{\rm 1} = 0.17374, \hskip1cm & c_{\rm 2} =-0.23047,\\
    c_{\rm 3} = 0.12149, \hskip1cm & c_{\rm 4} = 0.16940, \hskip1cm & c_{\rm 5} = -0.04509,\\
    c_{\rm 6} = 0.04509, \hskip1cm & c_{\rm 7} = 0.16658, \hskip1cm & c_{\rm 8} = 0.17511.\\
    \end{array}
\label{4qubits}
\end{equation}

\noindent
Note that Hamiltonian in Eq. \ref{4qubits} 
commutes with $Z_1$ and with $Z_3$. Therefore, the Hamiltonian is block-diagonal with $4$ blocks, each corresponding to a particular computational basis setting of the qubits $1$ and $3$.
This can be used to find a Hamiltonian with the same ground energy expressed in $2$-qubit space  \cite{qiskit.chemistry}:

\begin{equation}
	\hat{H}_{\mathrm{H_2}} 
	=  c_{\mathrm 0} \; \boldsymbol{1} + c_{\mathrm 1} \; Z_{\mathrm 0} 
	+ c_{\mathrm 1} \; Z_{\mathrm 1} 
	+ c_{\mathrm 2}  \; Z_{\mathrm 1}Z_{\mathrm 0}
	+ c_{\mathrm 3}  \; X_{\mathrm 1}X_{\mathrm 0},
    \label{H2_2_qubits}
\end{equation}

\noindent
with the coefficients

\begin{equation}
    \begin{array}{ll}
    c_{\rm 0} = -1.05016, \hskip1cm & c_{\rm 1} = 0.40421,\\
    c_{\rm 2} = 0.01135, \hskip1cm & c_{\rm 3} = 0.18038.\\
    \end{array}
    \label{2qubits}
\end{equation}	

\noindent With two equivalent formulations (both Hamiltonians have the same ground state energy) of the problem in hand, we will be able to investigate which form is more amenable to VQE optimization. They give us a chance to explore various problem sizes, parametrized state ansatzes, energy landscapes, and investigate how noise affects the procedures, as the optimizations will involve different numbers of gates.

We are now ready to run the VQE algorithm. Its input is a Hamiltonian expressed in a qubit basis and its aim is to find the eigenvector with the lowest eigenvalue. For this, the VQE iterates these four steps:

\begin{enumerate}
	\item {(quantum)} prepare a parametrized quantum state on a quantum device,
	\item {(quantum)} measure each Hamiltonian term (requires repetitions of step 1),
	\item {(classical)} sum the expectation values of the Hamiltonian terms to estimate the energy of the parametrized state,
	\item {(classical)} use the energy value to update the parameters of the trial quantum state,
\end{enumerate}
until we meet the convergence criteria of the classical optimization method.

To minimize the expectation value of the energy, we choose 
the simultaneous perturbation stochastic approximation (SPSA)~\cite{SPSA,Spall1, Spall}, classical optimization method implemented in {\em qiskit}. SPSA is a pseudo-gradient method for optimizing problems with varying numbers of unknown parameters. 
SPSA starts with the initial vector of parameters $\vec{\theta}^{}_{\mathrm{0}}$. In each iteration, the parameter vector is simultaneously shifted twice as 

\begin{equation}
    \vec{\theta}^{\pm}_{\mathrm{k}} = \vec{\theta}^{}_{\mathrm{k}} \pm c^{}_{\mathrm{k}}\vec{\Delta}^{}_{\mathrm{k}},
\end{equation}

\noindent
where $c^{}_{\mathrm{k}}$ is a preassigned positive number, ${\mathrm{k}}$ is the iteration number and $\vec{\Delta}^{}_{\mathrm{k}}$ is a randomly generated vector (from the Bernoulli distribution). To approximate the gradient at $\vec{\theta}^{}_{\mathrm{k}}$, we utilize the gradients at $\vec{\theta}^{+}_{\mathrm{k}}$ and $\vec{\theta}^{-}_{\mathrm{k}}$ as

\begin{equation}
    \vec{g^{}}_{\mathrm{k}}(\vec{\theta}^{}_{\mathrm{k}}) = \frac{\bra{\psi(\vec{\theta}^{+}_{\mathrm{k}})} H \ket{\psi (\vec{\theta}^{+}_{\mathrm{k}})} - \bra{\psi(\vec{\theta}^{-}_{\mathrm{k}})} H \ket{\psi (\vec{\theta}^{-}_{\mathrm{k}})}}{2c^{}_{\mathrm{k}}}\vec{\Delta}^{}_{\mathrm{k}}.
    \label{SPSAgrad}
\end{equation}

\noindent
In each iteration step we thus need to measure the energies of two quantum states, prepared with parameter settings  $\vec{\theta}^{+}_{\mathrm{k}}$ and $\vec{\theta}^{-}_{\mathrm{k}}$. 
We then update the underlying parameters $\vec{\theta}^{}_{\mathrm{k}}$ to  

\begin{equation}
    \vec{\theta}^{}_{\mathrm{k} + 1} = \vec{\theta}^{}_{\mathrm{k}} - a^{}_{\mathrm{k}}\vec{g}^{}_{\mathrm{k}}(\vec{\theta^{}_{\mathrm{k}}}),
\end{equation}

\noindent
where $a^{}_{\mathrm{k}}$ is a preassigned positive number and $\vec{g}^{}_{\mathrm{k}}(\vec{\theta}^{}_{\mathrm{k}})$ is the approximated gradient \eqref{SPSAgrad} that depends on $\vec{\theta}^{+}_{\mathrm{k}}$ and $\vec{\theta}^{-}_{\mathrm{k}}$. 

The optimization procedure runs for a number of iterations controlled by the \texttt{maxiter} parameter (in {\em qiskit}’s implementation). We then update the parameter vector $\vec{\theta}^{}_{\mathrm{k}}$, and measure the final value of energy for the underlying optimized parameters.
There is a subtlety that influences the total number of function evaluations. {\em Qiskit}'s implementation of SPSA includes additional initial exploration -- a calibration phase that depends on the maximum number of iterations with $\textrm{min}\left\{\texttt{maxiter}/5,\,25\right\}$ steps. 



There are many possible choices for the quantum circuit that prepares the trial state for simulating the molecules using VQE. We employ 
the hardware-inspired (easy to implement) $R_{\mathrm{y}}R_{\mathrm{z}}$ and $R_{\mathrm{y}}$ variational forms, accompanied by linear entanglement \cite{forms}.
The circuits consist of two main layers: (i) a layer of parametrized $R_{\mathrm{y}}$ or $R_{\mathrm{y}}$ and $R_{\mathrm{z}}$ single qubit rotations applied on each qubit in the quantum register, alternating with (ii) an entanglement-creating layer of CNOT gates. We can easily alternate these two layers, creating circuits of varying depth, increasing the complexity of the ansatz, and the number of parameters to be optimized.

\begin{figure}[ht]
	\centering
	\includegraphics[width=1\linewidth]{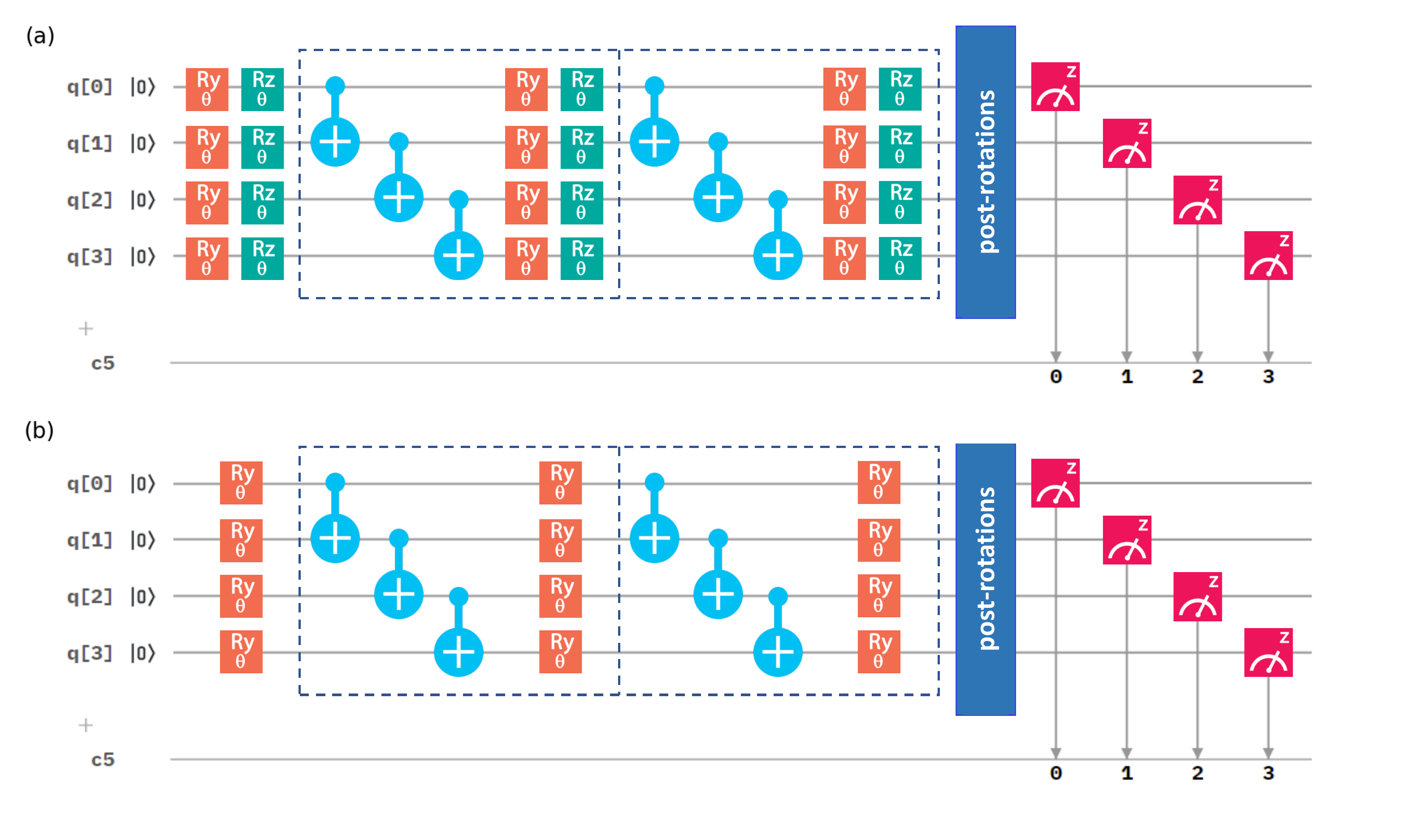}
	\caption{A scheme of the 4-qubit quantum circuits for calculations of the ground state energy. The circuit depth was set to 2. Dashed lines represent blocks of two layers -- the entangled layer and the rotation layer. (\textbf{a}) The $R_{\mathrm{y}}R_{\mathrm{z}}$ variational circuit. (\textbf{b}) The $R_{\mathrm{y}}$~variational circuit.}
	\label{fig:circuit_4}
\end{figure}

The simplest variational circuit of depth $d = 1$ consists of just the entanglement-creating layer, nested between two parametrized rotation layers. We draw the 4-qubit variational circuits of depth $d=2$ in Figure~\ref{fig:circuit_4}. 
For $q$ qubits and depth $d$, the $R_{\mathrm{y}}R_{\mathrm{z}}$ variational circuit has $2 q (d + 1)$ parameters -- single qubit rotations, while the $R_{\mathrm{y}}$ variational circuit has $q (d + 1)$ of them. 

Our qubit Hamiltonians \eqref{H2_4_qubits} and \eqref{H2_2_qubits} contain terms with the Pauli $X$ operators. To estimate their expectation values, we need to measure them in a non-diagonal basis.
As only computational-basis measurements are available on the quantum processors, we need to
utilize basis-switching single-qubit gates (post-rotations). For the Pauli operator $X$, we thus use a $\pi/2$ rotation around the $y$ axis, performed by gate $R_{\mathrm{y}}(\pi/2)$, and subsequently measure in the computational ($Z$-)basis. 

Let us note that we do not need to estimate all of the Pauli terms in our Hamiltonians individually. We can save resources and reduce the number of required measurements and state preparations by grouping the Pauli operators that require the same post-rotations in the tensor product basis sets \cite{kandala}.

To access the real quantum devices, we use the publicly available cloud-based quantum computing platform IBM Quantum~\cite{IBM}. We find that access to real quantum processors is still limited for a larger systematic study of different types of errors and noise. Therefore, we have heavily relied on classical simulations of quantum processors. The simulations were performed both in an ideal noise-free manner and including noise. Importantly, we have used the noise-related parameters of the IBM quantum processors, which are publicly available (and changing) on a daily basis. Thereby, our simulations 
could most closely mimic the actual gates and computations executed on a real device. 

The published noise-related data include an approximate noise model consisting of
(i)~single-qubit and two-qubit gate errors -- depolarizing errors followed by thermal-relaxation errors, (ii)~single-qubit readout errors on all measurements, and (iii)~all errors including gate, readout and thermal-relaxation errors. 
 Since calibration data changes frequently, in noisy simulation results we used the error model from the same calibration for most of the trials.

Finally, in some tests involving real quantum hardware (see Figure~\ref{fig:bogota}) we employ a simple error-mitigation method. It uses least-squares fitting to obtain the error-mitigated outcome probabilities by using a calibration matrix \cite{ErrMit}. 

\section{Results and discussion}

There are several possible sources of error for the energies calculated by the VQE. 
First, {\em statistical errors} in intermediate and final energy estimations, caused by the probabilistic nature of quantum mechanics. 
Second, {\em Hamiltonian representation and state-preparation ansatz} errors, caused by approximations in the Hamiltonian (restricted basis set), as well as the space of states we search over, given by our parametrized quantum circuit.
Third, {\em hardware errors} present in noisy quantum devices running the quantum part of VQE. We study these using simulators of noisy devices, as well as real quantum processor runs.
We will now investigate the influence of these errors, and discuss ways and the costs of mitigating them. 
We will do this for the H$_2$ molecular Hamiltonians from the previous Section.

Note that there are several factors which limit the number of gates/circuit repetitions we can execute.  First, we have only limited access to the quantum processors, which severely restricts the number of shots we can use for each datapoint. Second, {we do not have unlimited time either}. Third, the gates themselves are noisy, so increasing system sizes or gate numbers won't work without serious error mitigation. 
Thus, simply increasing the number of repetitions/circuit size is not a solution to these errors, and we must work harder on error mitigation to obtain trustworthy results -- molecular energies with chemical precision.

\subsection*{Number of quantum computer calls}
While searching for optimal parameters, the VQE optimization requires many calls to a quantum subroutine: prepare a candidate state and measure its energy. Even if we imagined having a noiseless quantum processor, its probabilistic nature and impossibility of measuring noncommuting observables simultaneously would force us to rely on averages over several measurements, which carry stochastic noise. 
We thus start by studying how the number of circuit (gate) evaluations on a quantum processor influences the convergence of the VQE algorithm, on ideal quantum devices.

Our goal is to understand and efficiently fight the influence of stochastic errors, inherent in VQE regardless of the quantum hardware. We want to optimally utilize a limited number of gate executions.
For this, we have two options. First, we can increase the number of shots (controlled by the \texttt{shots} parameter in the programs) for each evaluation of readout probabilities. Second, we can increase the maximum number of iterations (the parameter \texttt{maxiter}) allowed for the classical optimization subroutine.

Increasing the number of shots, improves the precision of the measurement outcomes for each of the terms in \eqref{H2_2_qubits} or \eqref{H2_4_qubits}. This makes the energy function more stable and easier for the classical optimization routine to minimize. 
Meanwhile, increasing the allowed number of iterations for the classical optimizer provides it with a better chance to get out of existing local minima and/or to fine-tune the final output value, but again comes at the cost of increasing the number of times quantum computer needs to be accessed.
Intuitively, it can be expected that increasing the number of quantum gate executions in either case increases the precision of the output. 
However, presently the number of gate executions on a quantum computer is a limiting factor for most users -- access to real quantum processors is either restricted or quite costly \cite{ShotPrice}.
Additionally, an excessive number of quantum gate executions significantly impacts the real running time of the algorithm.

\begin{figure}[ht]
	\includegraphics[width=1\linewidth]{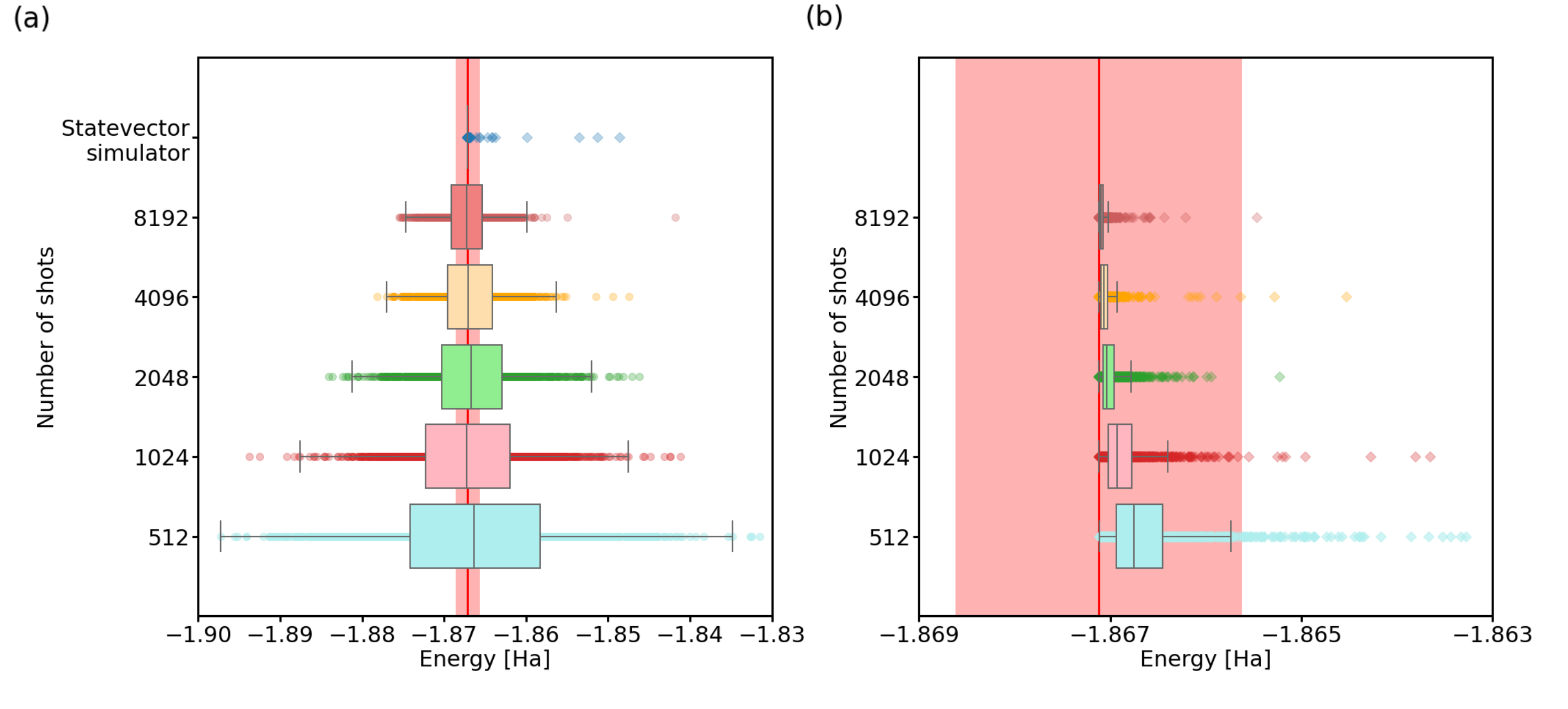}
	\caption{
	VQE energy estimates improvement with number of shots. We calculate the ground state energy of H$_{\mathrm2}$, using the VQE on a two-qubit system, the SPSA optimization algorithm with unrestricted maximum interations, and the $R_{\mathrm{y}}$ variational circuit accompanied by linear entan\-gle\-ment. The red line represents the physical ground state energy ($-1.86712$ Hartree) and the light-red background represents the chemical accuracy regime ($\pm 0.0015$ Hartree). (\textbf{a}) A boxplot of VQE results, with each box representing $1000$ independent runs of the algorithm. (\textbf{b}) A statevector simulator calculation of the energies for the $1000$ result states from (a). Note that all the energies are obtained above the real ground state energy.
	\label{fig:shots}
	}
\end{figure}

\begin{figure}[ht]
	\includegraphics[width=1\linewidth]{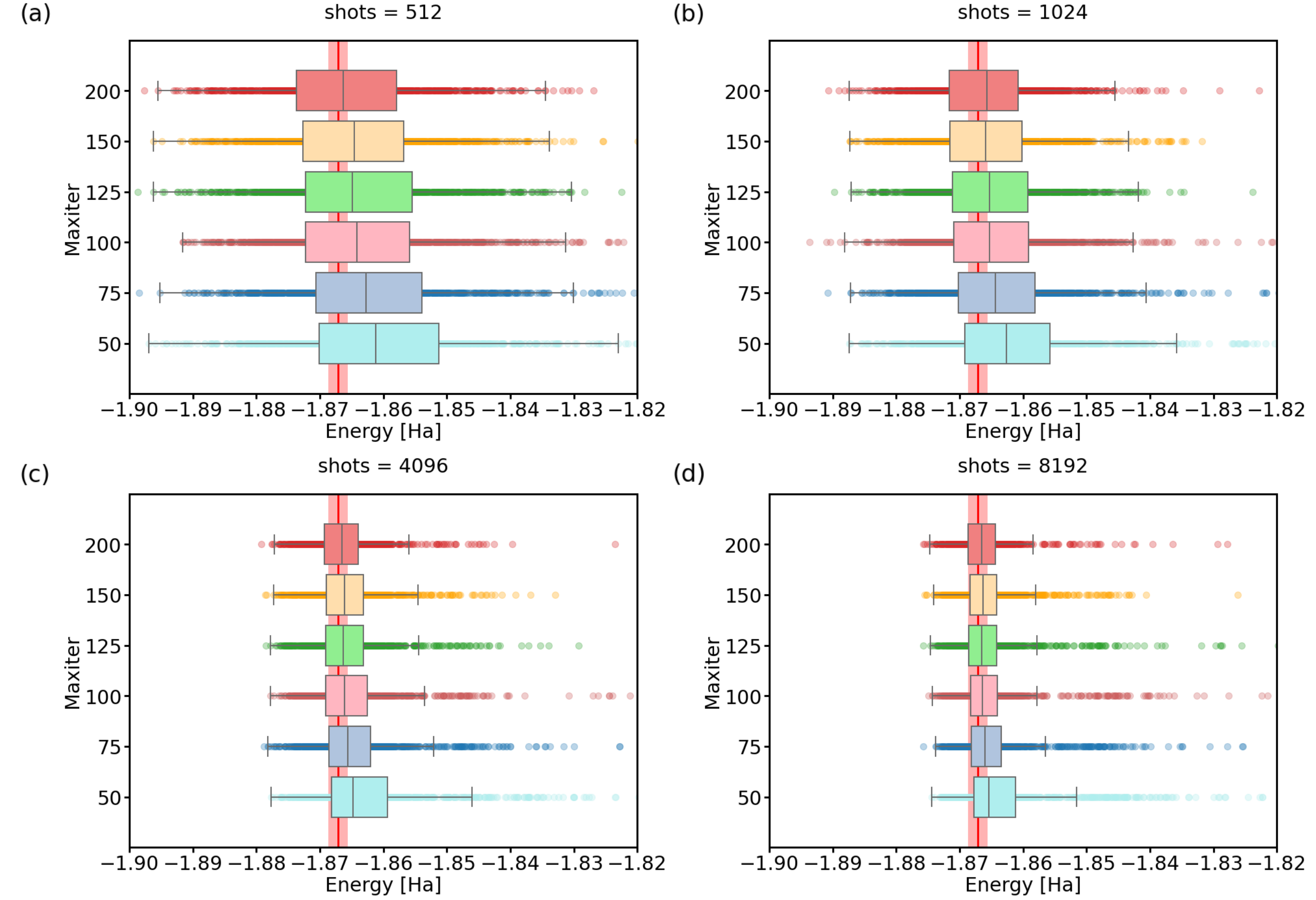}
	\caption{ (\textbf{a})-(\textbf{d})  A boxplot visualization of $1000$ optimized ground state energies for various maximum number of iterations of the~SPSA optimizer. We performed calculations using the $R_{\mathrm{y}}$ variational form. The red line represents the physical ground state energy ($-1.86712$ Hartree) and the light red background represents the chemical accuracy ($ \pm 0.0015$ Hartree). The  \texttt{shots} parameter was set to: (\textbf{a})~$512$, (\textbf{b})~$1024$, (\textbf{c})~$4096$, (\textbf{d})~$8192$.
	\label{fig:maxiter}
	}
\end{figure}

In this light, it is an interesting question to find the minimum number of quantum gate executions that  result in VQE output within chemical precision of the real energy value.
In our study, we ran the algorithm for different settings of \texttt{shots} and \texttt{maxiter} parameters, on the two qubit hydrogen molecule Hamiltonian, using the noiseless quantum simulator. Each combination of the settings was run $1000$ times. We present the experimental results in detail in Table \ref{tab:QuantumUses} (see Appendix~\ref{AppendixDataTables}), and visualize it here in Figures~\ref{fig:shots}~and~\ref{fig:maxiter}.

We visualize the data using a boxplot. The box shows the middle two quartiles of the data (interquartile range, IQR), with a marked median. Outliers are determined using Tukey fences \cite{tukey77}: a distance of $1.5$ times the IQR in each direction, as in Figure~\ref{fig:shots}(a), or the extremes of our data, as in Figure~\ref{fig:shots}(b), where all of our computed energies fall above the red line, close to the box. 

From Figure \ref{fig:shots}(a), we can see that with an ideal quantum computer and unrestricted iterations of the SPSA algorithm, VQE performs increasingly better with the increase of \texttt{shots} parameter. 
The limit of infinitely many shots is simulated using a statevector simulator, which calculates the outcome probabilities directly from the laws of quantum mechanics.
It is important to note that when using a finite number of shots, the algorithm can also output energy values lower than the actual minimum energy. 
Thanks to stochastic noise, the estimates of outcome probabilities for the Pauli terms in \eqref{H2_2_qubits}, calculated from a limited number of shots, can be non-physical.
To obtain realistic final optimized energies, one should thus invest in precise final energy readout for the optimized state, using a larger number of shots.

This also rises another interesting question: can the inaccuracy in outcome probabilities be overcome by the SPSA algorithm? 
In other words, are the states discovered by SPSA actually close to the minimum eigenvector of the Hamiltonian, and {is most of the energy spread seen in the results caused} by the small number of shots?
We recalculated the energies of the discovered states in Figure \ref{fig:shots}(a), using the statevector simulator.
The results (see Figure \ref{fig:shots}(b)) somewhat surprisingly show that indeed, most of the discovered states have real energies within the chemical accuracy. 

In the context of trying to minimize quantum computer calls, this suggests that the VQE algorithm can be run with a small number of shots during the run of the classical optimization function 
and after the optimization produces the final result, the energy of the candidate state with minimum energy needs to be recalculated once more, using a much larger number of shots.
However, for this to work, we have to use a classical optimizer which can deal well with noisy data. We could use Bayesian methods \cite{2017arXiv170607094L,2018arXiv180702811F}, 
which could be quite slow, needing to invert large matrices to guess next points for search. Or we could use other optimization methods like NEWUOA 
\cite{Powell2006}, which can be adapted to work on noisy data.
Note that in initial stages of this work, we have used the COBYLA optimization method~\cite{cobyla}, and concluded that very large numbers of shots were necessary for the method to converge at all.
As described in the Methods Section, in the end we have decided on the use of SPSA, thanks to its natural noise resilience, as it explores random points around the current parameter vectors.

On the other hand, at least for our selected Hamiltonian, increasing the \texttt{maxiter} parameter and waiting longer for convergence seems to have diminishing returns after surpassing a value of approximately $\texttt{maxiter} = 100$.
We confirm this in detail in in Figure~\ref{fig:maxiter}. There, we simulate $1000$ VQE calculations for various combinations of \texttt{maxiter}/\texttt{shots} settings. For the same value of \texttt{shots} parameter, increasing the \texttt{maxiter} parameter only decreases the number of outliers, but the spread of the data remains almost unchanged. 
Both of these observations suggest that only a small fraction of runs can benefit from the increase of \texttt{maxiter} parameter beyond $100$.


\begin{figure}[ht].
	\includegraphics[width=1\linewidth]{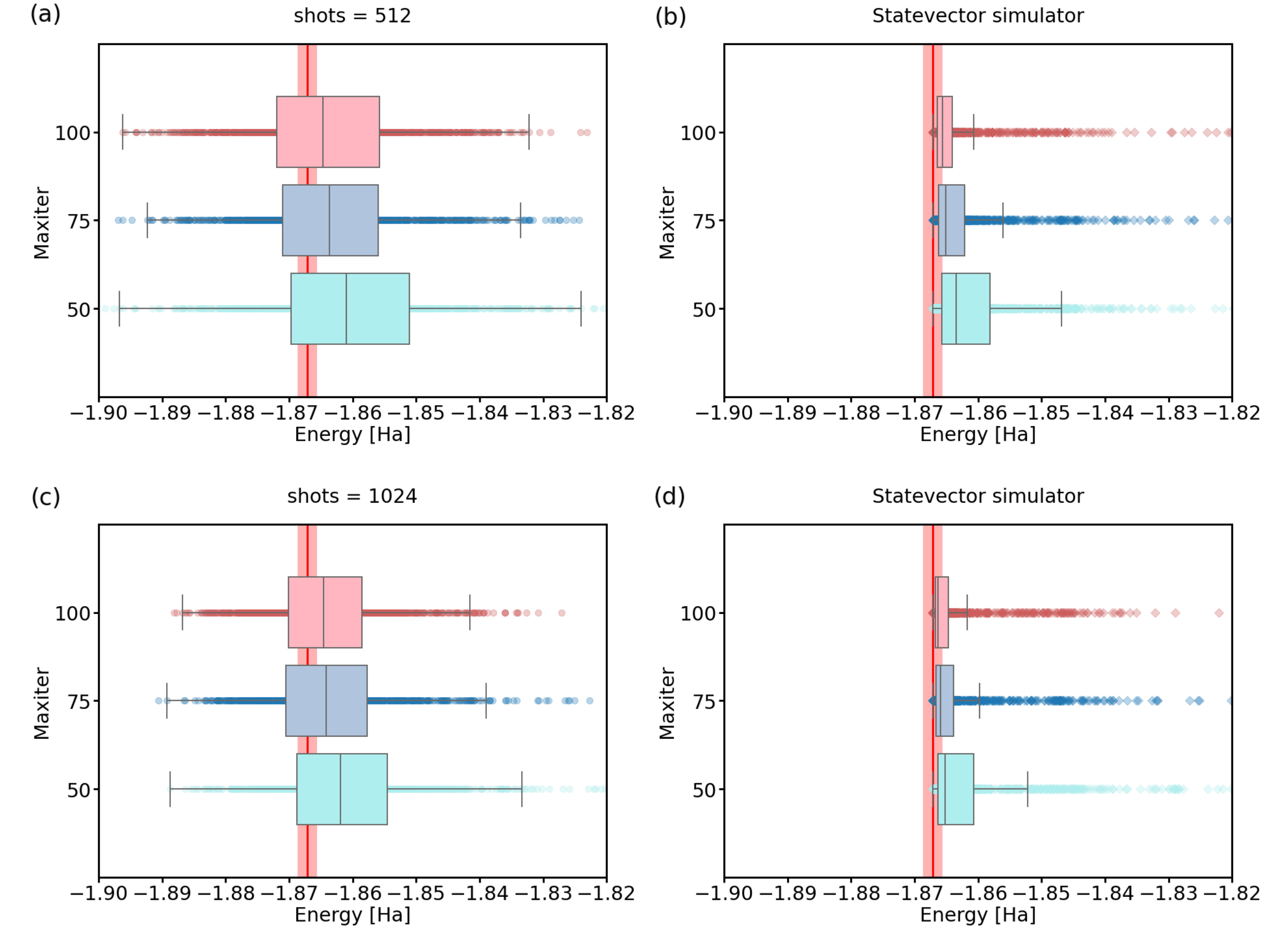}
	\caption{ 
	\label{fig:maxiter2} A recalculation of energies using statevector simulator for experiments with \texttt{maxiter} $\in \{50,75,100\}$, and (\textbf{a}), (\textbf{b}) \texttt{shots}=$512$ ; (\textbf{c}), (\textbf{d}) \texttt{shots}=$1024$. The red line represents the physical ground state energy ($-1.86712$ Hartree) and the light red background represents the chemical accuracy ($ \pm 0.0015$ Hartree).
	}
\end{figure}

In order to evaluate the energies more precisely after the last step, we recalculated energies using state vector simulator for \texttt{shots} parameter set to $512$ and $1024$, with $\texttt{maxiter}$ settings $50,75$ and $100$. Results of this calculation can be found in Fig.\ref{fig:maxiter2} and they suggest that the combination of \texttt{shots} = $1024$ and \texttt{maxiter} = $75$ already provides median energy in the chemical accuracy for the $2$-qubit H$_2$ Hamiltonian. 

\subsection*{Choice of Hamiltonian and state-preparation ansatz}

In this Section we study the influence of the size and form of the quantum circuit used in the VQE calculation. 
The most straightforward way to reduce the complexity of the VQE algorithm is to find the smallest possible qubit representation of the studied Hamiltonian.
Indeed, much effort today is focused on efficient encodings of fermionic Hamiltonians
(see e.g.\cite{PhysRevB.104.035118,PRXQuantum.2.030305} and references therein), as getting rid of extra qubits (and operations on them) decreases noise and can avoid space limitations.
To highlight the importance of finding minimal representations, we have chosen two equivalent variants of H$_{\mathrm2}$ molecule Hamiltonian, one using $2$ and another one using $4$ qubits (see the Methods section). 

Once we have chosen a Hamiltonian and its qubit implementation, VQE needs a choice of the variational circuit for state-preparation. We have a vast choice of parametrized gates, arrayed in multiple rounds. This results in varying numbers of classical parameters to be optimized, as well as different reachable quantum states (ansatz quality), with varying amounts of entanglement \cite{Cerezo2021,Funcke2021dimensional,Nakaji2021expressibilityof}.
Here, we focus on two types of variational circuits -- the $R_{\mathrm{y}}$ and the $R_{\mathrm{y}}R_{\mathrm{z}}$ linear forms with a varying number of rounds, shown in Figure~\ref{fig:circuit_4} for $4$ qubits (the $2$-qubit circuits are obvious simplifications). 

Note that all the results in this subsection were obtained using a noiseless quantum simulation, in order to highlight the influence of the Hamiltonian/ansatz choice in the ideal case. 
Of course, once we consider noise and statistical readout errors, the Hamiltonian/circuit choice translates to more errors, thanks to wider circuits with a larger number of gates.

\begin{figure}[ht]
	\includegraphics[width=1\linewidth]{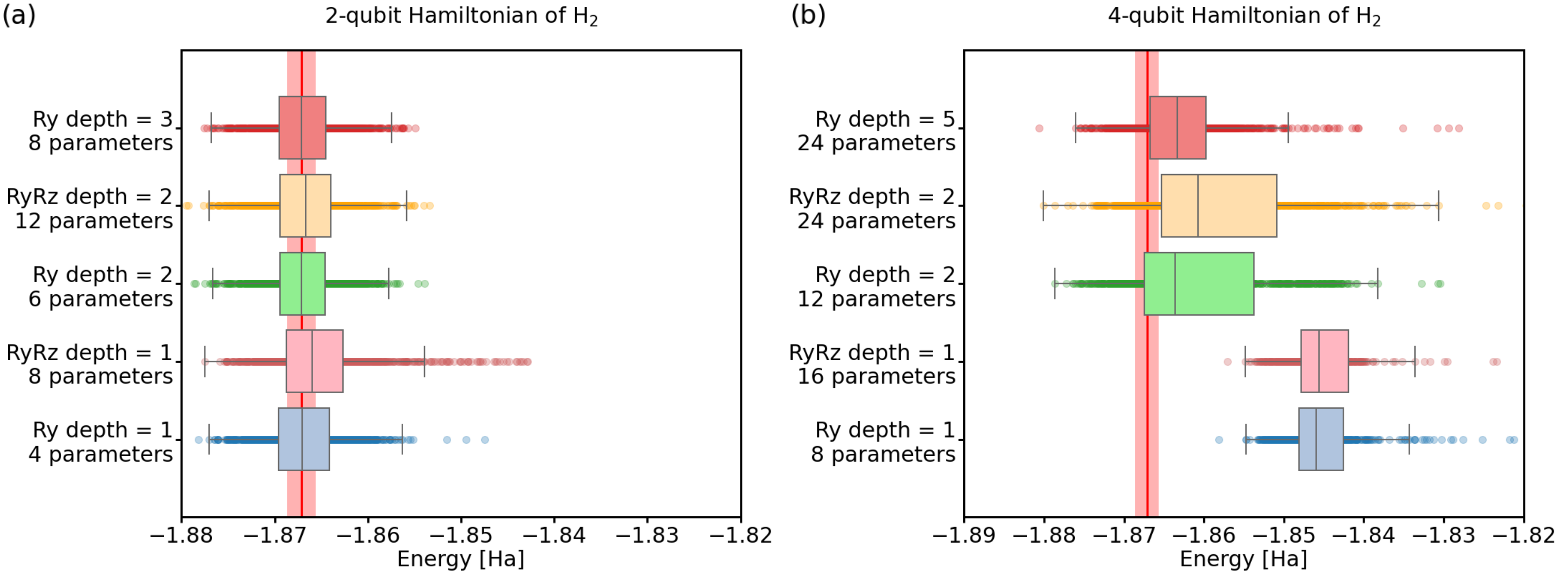}
	\caption{Comparison of the VQE energies for H$_2$ using the $R_{\mathrm{y}}$ and $R_{\mathrm{y}}R_{\mathrm{z}}$ variational circuit forms. We used a noiseless quantum simulator and $4096$ shots; (\textbf{a}) $2$-qubit Hamiltonian. (\textbf{b}) $4$-qubit Hamiltonian. The red line represents the physical ground state energy ($-1.86712$ Hartree) and the light red background represents the chemical accuracy ($ \pm 0.0015$ Hartree).
	}
	\label{fig:forms}
\end{figure}

In Figure \ref{fig:forms}, we depict our results for the convergence of the VQE algorithm, depending on the Hamiltonian choice ($2$- or $4$-qubit), and the ansatz complexity (type and number of rounds). 

The two-qubit Hamiltonian \eqref{H2_2_qubits} turns out to be simple enough that the depth of the variational ansatz (number of layers) does not significantly influence the final optimized energies.
Moreover, we can see that the simpler $R_{\mathrm{y}}$ ansatz outperforms $R_{\mathrm{y}}R_{\mathrm{z}}$ ansatz. 
This confirms the intuition that the simplest ansatz containing the solution will also have the best performance,
thanks to solution space having less parameters to optimize over.
Additionally in noisy regime, simpler ansatz requires less gates for implementation, leading to further advantage.

On the other hand, for the $4$-qubit Hamiltonian \eqref{H2_4_qubits}, both $R_{\mathrm{y}}$ and $R_{\mathrm{y}}R_{\mathrm{z}}$ variational forms with $1$ layer are not expressive enough and the search space does not contain the state with the lowest energy.
Using two layers already alleviates this problem. 
Again, we see that the $R_{\mathrm{y}}$ form performs better. 
Using a $12$-parameter, two-level $R_{\mathrm{y}}$ ansatz achieves better convergence than the $24$-parameter, two-level $R_{\mathrm{y}}R_{\mathrm{z}}$ ansatz.
Moreover, if we allow the $R_{\mathrm{y}}$ form with $24$ parameters as well, the circuit has depth $5$ and outperforms the $R_{\mathrm{y}}R_{\mathrm{z}}$ ansatz even more significantly.
This can be partially explained by the fact that for the same number of search space parameters, the $R_{\mathrm{y}}$ ansatz can achieve more intricately entangled states, due to the increased number of entanglement layers.

These results only underline the importance of an efficient choice of Hamiltonian encoding and ansatz parametrization. 
More qubits mean the need for more intricate ansatzes, while introducing more statistical errors when estimating the energies of individual terms, and thus affecting the overall readout precision, even in an ideal (noiseless) case. {Further, more complex Hamiltonians can cause the optimization algorithm to converge to a local minimum, thus causing a systematic error. This is in more detail treated in Appendix \ref{app:convergenceLocMin}.}


\subsection*{Imperfect quantum devices}
So far, we studied the convergence of the VQE algorithm assuming a perfect quantum computer available for the quantum subroutines.
This analysis was possible thanks to relative simplicity of classically simulating the few-qubit quantum circuits used during our VQE algorithm runs. 
We wish to eventually scale up the calculations, and obtain results from the hybrid classical-quantum VQE which are not available classically. For this, we will have to rely on real quantum devices.
Of course, current quantum processors are far from perfect, so the calculations will be inherently noisy, 
influencing also the convergence of VQE algorithm.

Let us group the possible errors into two basic types -- gate errors and readout errors. 
This is a natural divide, since the number of measurement (readout) errors scales only with the number of qubits and can be mitigated with classical techniques. Meanwhile, the gate errors inevitably scale with both depth of the computational ansatz and the number of qubits, unless we use error correction/mitigation techniques.

\begin{figure}[ht!]
	\centering
	\includegraphics[width=1\linewidth]{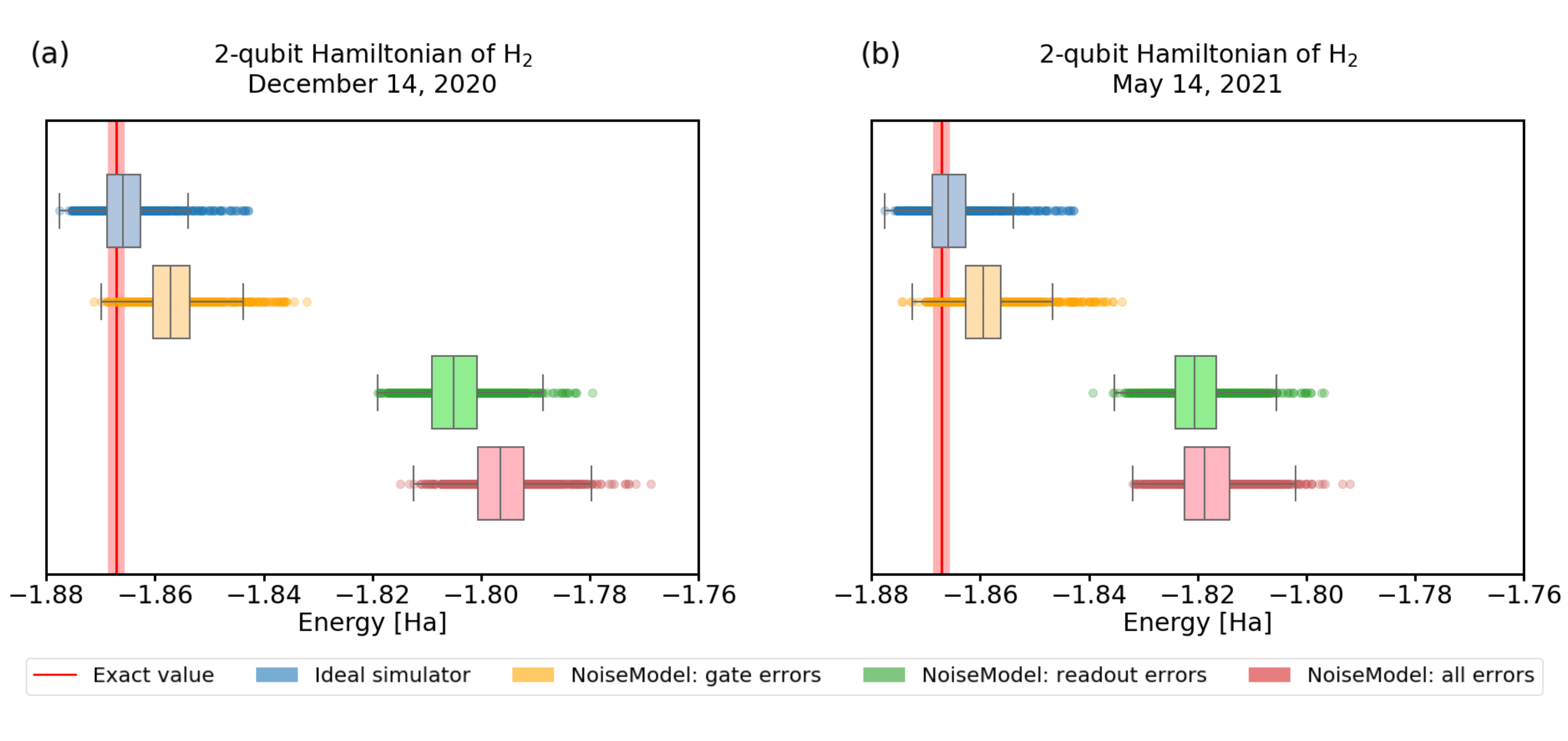}
	\caption{A comparison of VQE results with noise calibration from two different dates. We used quantum backend {\em ibmq$\_$santiago}. All the other parameters in the simulation were kept the same, using $4096$ shots and the $R_{\mathrm{y}}R_{\mathrm{z}}$ variational ansatz. The red line represents the physical ground state energy ($-1.86712$ Hartree) and the light red background represents the chemical accuracy ($ \pm 0.0015$ Hartree).}
	\label{fig:12}
\end{figure}

\begin{figure}[ht]
	\centering
	\includegraphics[width=1\linewidth]{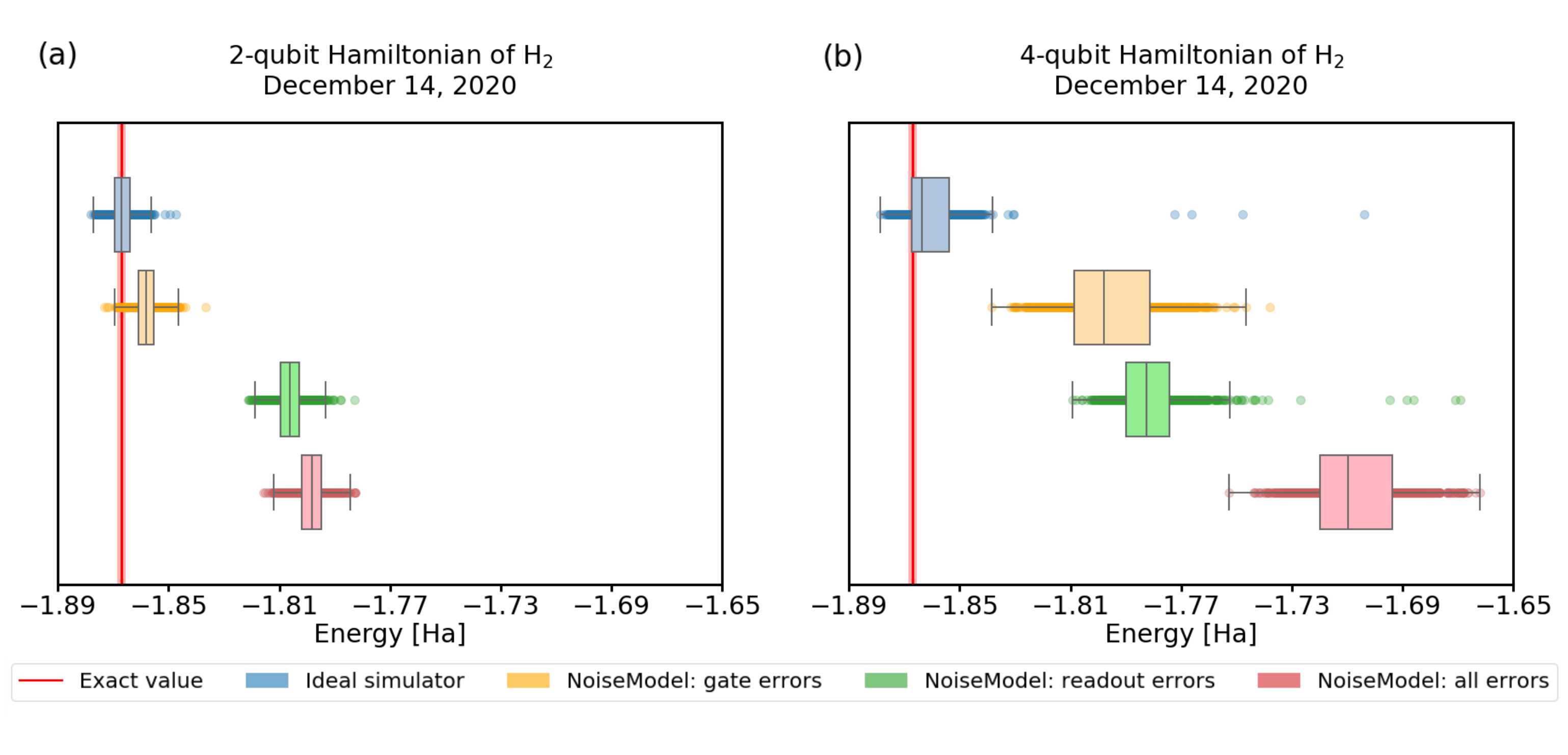}
	\caption{The calculations of H$_{\mathrm2}$ energy were performed using simulator of real quantum hardware using a $2$ and a $4$ qubit Hamiltonian. We used quantum processor {\em ibmq$\_$santiago} with noise model from December $14$ $2020$. The number of shots was set to $4096$. We minimized energy with the SPSA optimizer: (a) in case of $2$-qubit Hamiltonian we used unrestricted maximum number of iterations and the $R_{\mathrm{y}}$ variational form with depth $1$, (b) in case of $4$-qubit Hamiltonian the maximum number of iterations was set to $400$ and we used the $R_{\mathrm{y}}$ form with depth $2$. The red line represents the physical ground state energy ($-1.86712$ Hartree) and the light red background represents the chemical accuracy ($ \pm 0.0015$ Hartree).}
	\label{fig:errors}
\end{figure}

In Figures \ref{fig:12} and \ref{fig:errors}, we show the convergence of the VQE using simulations of noisy quantum processors. To obtain realistic parameters for the simulation, we used the error characterization of the real IBM devices, obtained directly from {\em qiskit}.

Our first observation is that the quality of calibration of recent experimental quantum processors changes frequently. Not only that, it significantly influences the convergence, as we demonstrate in Figure~\ref{fig:12}. There, we visualize the convergence of $1000$ trials with the same settings, using  calibration data from different dates{, showing results differing by $10\%$}.

Second, we observe that for these small circuits, the effect of measurement error is more significant than the gate errors. 
As we study this further in Figure \ref{fig:errors} for the $4$-qubit Hamiltonian \eqref{H2_4_qubits}, we find that readout errors still affect the convergence more, but the effect is less pronounced.
In Figure~\ref{fig:errors}(a) we use the $2$-qubit, $1$-round $R_y$ variational form, which requires $4$ single-qubit gates + $1$ CNOT, and up to $2$ Pauli operators for readout.
For Figure~\ref{fig:errors}(b), we recall results presented in Figure~\ref{fig:forms}, and choose the $4$-qubit, $2$-round $R_y$ variational form, which requires $12$ single-qubit gates + $6$ CNOTs, and up to $2$ Pauli operators for readout. 
The number of gates (circuit width $\times$ depth), and thus gate errors, grows faster than the number of estimation terms, which still scales with the circuit width. Thus, we expect that the effect of readout error will become less prominent for larger Hamiltonians compared to the gate errors.
Moreover, we can also implement various readout error mitigating strategies -- e.g. avoiding systematic bias by using bit flips before the final measurements.

\subsection*{Convergence of VQE on real quantum hardware}
\begin{figure}[h!]
	\centering
	\includegraphics[width=0.7\linewidth]{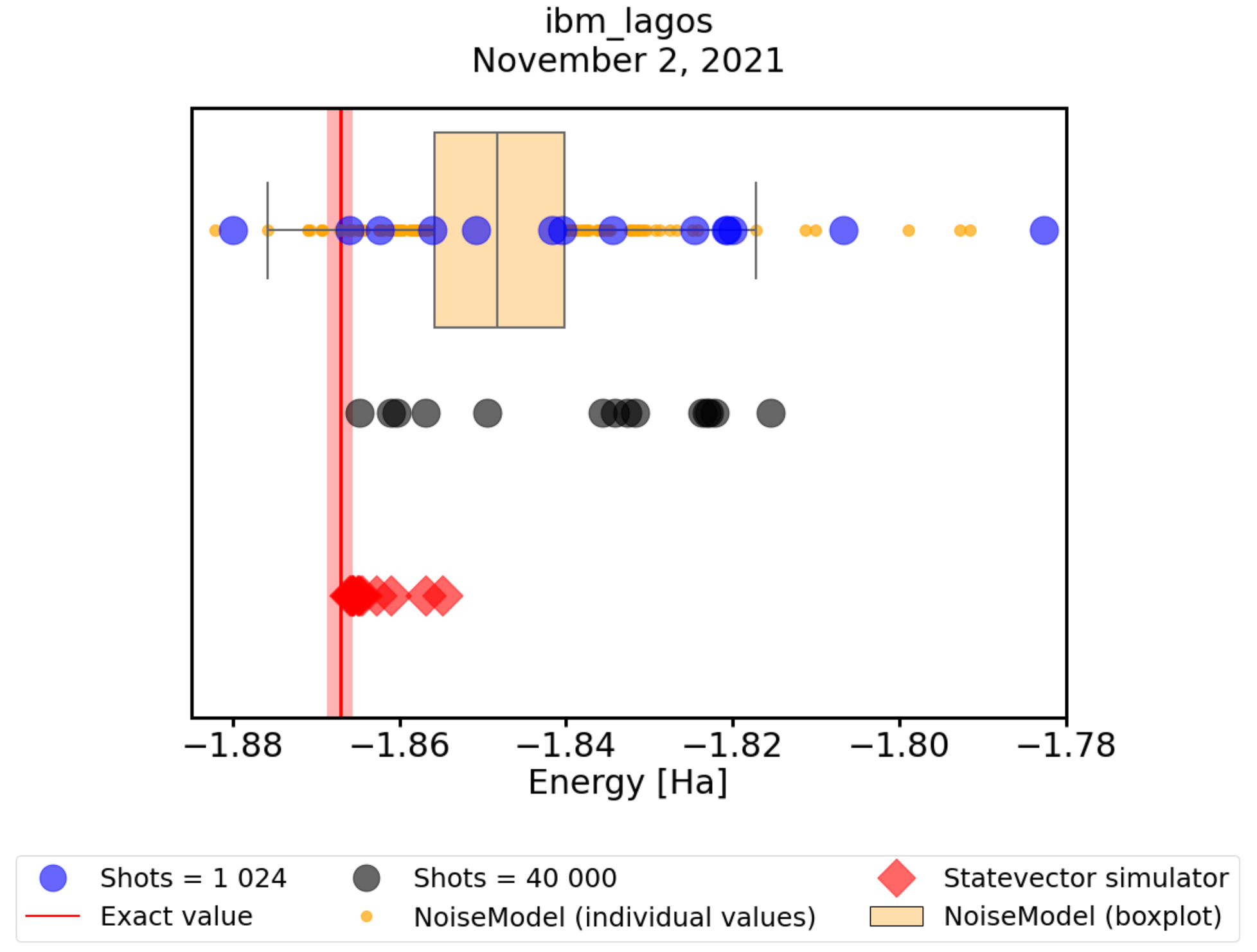}
	\caption{
	A comparison of results obtained with a noisy quantum simulator with error mitigation (orange points and box plot), with real runs of the {\em ibm$\_$lagos} quantum processor (blue points).
	The energies of real runs were subsequently recalculated with greater precission using $40000$ shots of the {\em ibm$\_$lagos} quantum processor (grey points). Finally, we recalculate their energies also with a statevector simulator (red diamonds).
	The red line represents the physical ground state energy ($-1.86712$ Hartree) and the light red background represents the chemical accuracy ($ \pm 0.0015$ Hartree).
	}
	\label{fig:bogota}
\end{figure}

Finally, in this Subsection we look at how the convergence predicted by the noisy simulator corresponds to convergence on real quantum hardware. 
For this, we performed an experimental runs on the {\em ibm$\_$lagos} quantum processor, with \texttt{maxiter} = $75$ and \texttt{shots} = $1024$, and compared it to noisy simulation results with the same parameter settings.
In order to mitigate the readout errors we performed a simple mitigation routine described at the beginning of this Section.
In Figure~\ref{fig:bogota}, we observe that in fact the predictions agree fairly well with practice.
Further, we tested our hypothesis that the discovered states are very close to optimal ones, but their energies have to be evaluated more precisely.
To this end we calculated their energies using $40000$ shots on the {\em ibm$\_$lagos} quantum processor ($40000$ shots result in $\approx 0.5\%$ error in energy calculations) as well as using a statevector simulator. 
In Figure~\ref{fig:bogota}, we can see that both of these techniques decrease the spread of the data points. However, only the statevector simulator consistently evaluates energies within the chemical accuracy from the exact ground state value.
This only shows that evaluating the energies using a large number of shots still suffers from hardware errors, which are not present when using the statevector simulator.
We conclude that a simple VQE implementation on real quantum hardware does not {\em consistently} find the minimum energy within the chemical accuracy.
Nevertheless, using both error-mitigation and precise energy readout one can obtain a trustworthy result: take the lowest of the precisely evaluated energies.

\section{Conclusions}


In this work we provided a guided tour of quantum chemistry calculations with the Variational Quantum Eigensolver on real quantum hardware. We analyzed the influence of different types of errors on the convergence of VQE. We divided them into several categories and studied them separately, in order to understand their impact and to find ways of avoiding them, while using our resources efficiently.
For this, we used noiseless and noisy classical simulations, as well as publicly available superconducting quantum processors.

The errors resulting from a probabilistic nature of quantum mechanics significantly differ from the hardware errors. While the statistical error increases the spread of VQE outcomes around the optimum, gate and readout errors not only influence the spread of the outcomes, but often prevent the VQE from finding the optimum altogether.

If we could use noiseless hardware, the stochastic errors could be mitigated by increasing the number of shots, and getting precise energy estimates. This turns out not to be that important in the intermediate steps of iterative algorithm, but crucial at the end. Even with smaller numbers of shots per iteration, our optimization was able to find near-optimal states. However, in order to rule out unphysical results (energies below the actual minimum energy), but also to avoid overestimating the energy, we must invest shots into the final, precise readout. We learned that this approach can significantly reduce the overall cost (gate evaluations) for VQE, important while access to real devices (and large runs) is limited, as well as time-consuming.

Second, it is crucial to search for efficient encodings and good trial-state ansatzes (hardware, and problem-motivated). It saves on both gate-evaluations, as well as on optimization complexity, reducing the number of required parameters to search over. We have demonstrated this in the comparison of the 4-qubit and 2-qubit Hamiltonians, and two types of ansatzes, with varying depth/number of parameters, for the same molecular hydrogen problem.


Finally, real devices' parametrization changes with time, and it is crucial to choose to run your experiments at times when they are calibrated well. Moreover, the current levels of noise are prohibitive, and add up so that the final results do not straightforwardly reach the desired chemical precision. In order to avoid this, error mitigation must be used, for the readouts (which contribute a surprising amount of error) as well as inside of the circuit.

\vspace{6pt} 



\authorcontributions{
{Conceptualization, I.M.,M.Pi., M.Pl., M.F., M.\v{S}.; 
methodology, M.Pi.,D.N.; 
software, I.M.; 
validation, I.M.,M.Pi., M.Pl.;
formal analysis, I.M.,M.Pi.;
investigation, I.M.,M.Pi., M.Pl., M.F.;
resources,  I.M.,M.Pi., M.Pl.;
data curation, I.M.;
writing---original draft preparation, I.M.,M.Pi., M.Pl., M.F. ; 
writing---review and editing, D.N.; 
visualization, I.M.; 
supervision, M.\v{S}., M.Pi., M.Pl.,M.F. ; 
project administration, M.\v{S} ; 
funding acquisition, M.\v{S}., M.Pi., M.Pl., M.F. 
All authors have read and agreed to the published version of the manuscript.}
}

\funding{
This research was funded by 
the Grant Agency of Masaryk University in Brno, Czech Republic, within an interdisciplinary research project No. MUNI/G/1596/2019 entitled "Development of algorithms for application of quantum computers in electronic-structure calculations in solid-state physics and chemistry". 
}

\conflictsofinterest{The authors declare no conflict of interest. The funders had no role in the design of the study; in the collection, analyses, or interpretation of data; in the writing of the manuscript, or in the decision to publish the~results.} 

\sampleavailability{No compounds were created in this study.}

\bigskip
\noindent\small{\textbf{Acknowledgements:}}
 We acknowledge the use of IBM Quantum services for this work. The views expressed are those of the authors, and do not reflect the official policy or position of IBM or the IBM Quantum team. We acknowledge the access to advanced services provided by the IBM Quantum Researchers Program. 
 We would like to thank Martin Saip for fruitful discussions.

\abbreviations{The following abbreviations are used in this manuscript:\\

\noindent 
\begin{tabular}{@{}ll}
VQE & Variational Quantum Eigensolver\\
STO-3G & Slater-type Orbital basis set with each orbital expanded into
       $3$ Gaussian functions\\ 
SPSA & Simultaneous Perturbation Stochastic Approximation
\end{tabular}}

\newpage
\reftitle{References}
\externalbibliography{yes}
\bibliography{sample}
\appendixtitles{yes} 
\appendixstart
\appendix
\newpage
\section{Convergence to local minima}\label{app:convergenceLocMin}


In this Appendix, we dig in beyond energy minimization and ask: are we finding the correct (ground) state? Our analysis above is primarily focused on identifying the impact of different types of noise and errors that are nowadays commonly occurring in physical realizations of superconducting quantum processors. 
We have purposely chosen the H$_2$ molecule as a material system of our case study because the ground state of H$_2$ is well known. 
In fact, it is a frequent textbook example in quantum chemistry and, therefore, it is rather straightforward to show the impact of noise. 
Importantly, there are two implicit underlying assumptions of our analysis. 
First, we expect  the minimization procedure within our methodology to aim solely at the ground state. 
Second, we assume that all the obtained values, covering a wide energy range, are indeed ground-state-related solutions that differ from the known exact ground-state energy only due to the noise.
These two aspects are mutually related and the validity of our assumptions sensitively depends on both the minimizing method and the studied Hamiltonian.
Had the noise-related range of energies been wider than the energy separation to the next higher-energy eigenvalue(s), and had this excited-state eigenvalue been found by the minimization procedure, yet another ``source of errors'' would have affected our analysis. 
This error type would be methodological, rather than noise-related. 
We address this issue below and show that the analysis focused on only the eigenvalues, i.e. energies, is insufficient.
Note that while estimating the final energy, we perform many measurements, and thus obtain a list of (significant) probabilities. 
This is not full tomography, which would be prohibitive for large scale systems. 
We only estimate the expectation values of Pauli operators present in the Hamiltonian.
In our case, we estimate probabilities for computational basis states in order to estimate expectation values of Pauli $Z$ operators, while estimating Pauli $X$ operators on two of the qubits forces us to do measurements in a rotated basis. 
We will now study how to utilize the resulting vectors of probabilities to distinguish different classes of the states produced by VQE.

\begin{figure}[ht]
	\centering
	\includegraphics[width=0.65\linewidth]{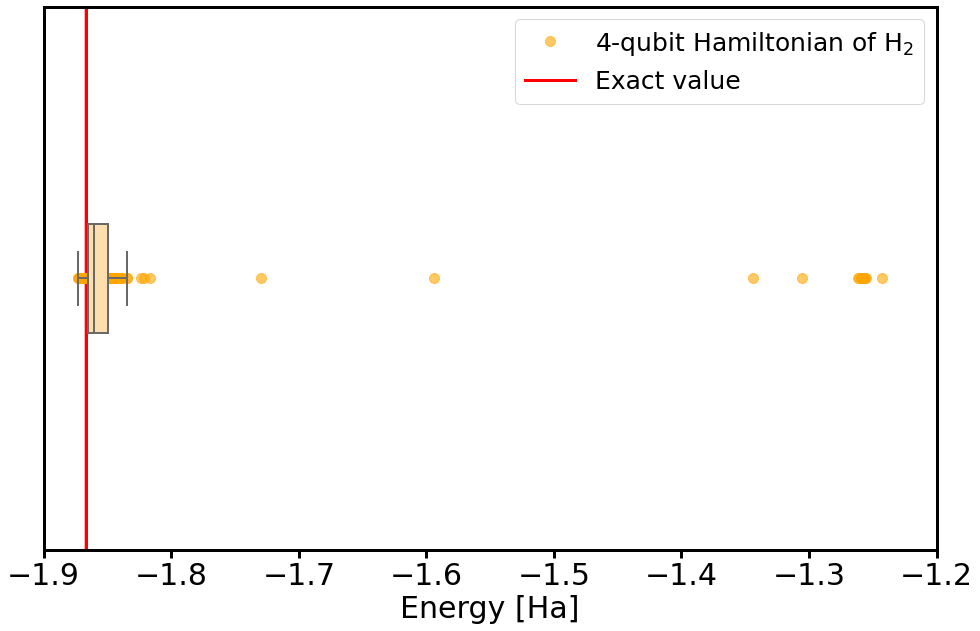}
	\caption{Ground state energies of the $H_2$ molecule from $200$ calculations on a noise-less simulator, using the $4$-qubit Hamiltonian \eqref{H2_4_qubits}.
	\label{fig:wide-range}}
\end{figure}

Figure~\ref{fig:wide-range} shows  our results for the $4$-qubit Hamiltonian \eqref{H2_4_qubits} in an energy range of $(-1.9, -1.1)$ Hartree. We can notice a set of clearly outlying values above the energy of $-1.3$ Hartree. 
They can be  identified in a very straightforward manner because they are separated from the main set of ground-state-related energies by quite a wide energy range that is free of other datapoints.

In order to discern what these outlying energies represent we next evaluate higher-energy eigenvalues of the
studied qubit Hamiltonians of H$_2$. 
The eigenvalues of the $4$-qubit Hamiltonian of H$_2$ molecule, as determined using classical techniques, are as follows: 
\begin{align*}
 (-1.867,-1.262, -1.262, -1.242, -1.242, -1.242, -1.160,\\ -1.160, -0.881, -0.465, -0.465, -0.341, -0.211, 0, 0.227).
\end{align*}

Their inspection reveals that while the ground-state energy of H$_2$ is very well separated from the other higher-energy eigenvalues, there are {several} eigenvalues in the energy range of $(-1.3, -1.1)$ Hartree, albeit some of them degenerated. 
A challenge that we face is to tell (i) those cases that are noise-related to the ground-state energy from (ii) those that are (again noise-related) to the other eigenvalues. 
In what follows we propose to distinguish between sets of values corresponding to different eigenvalues by applying a suitable similarity index to the measured probabilities of different basis states. 

Before we start discussing similarity measures, it is worth noting a few facts related to the measured probabilities of basis states. 
They are sets of {non-negative} real numbers smaller or equal to one. 
The number of elements in each set is equal to $n = 2^q$, where $q$ is the number of qubits, and the sum of all elements in each set of probabilities equals to one. 
As discussed above, the measured sets of probabilities are multiplied by values of integrals within the qubit Hamiltonian in order to evaluate the energies. 

Our analysis of similarities of vectors of measured probabilities below includes two different measures. 
The first one is  the Jaccard-Tanimoto (J-T) index, also known as the Jaccard similarity coefficient. 
It is used for gauging the similarity and diversity of sample sets. 
It was developed by Paul Jaccard~\cite{Jaccard} and independently formulated again by T. Tanimoto~\cite{Tanimoto}. 
The Jaccard-Tanimoto index of two sets X and Y is defined in general as the ratio of intersection of the two sets over their union
$
J-T(X,Y) = |X \cap Y| / |X \cup Y|. 
$
For two vectors $\{x_i\}, \{y_i\}$ with all components non-negative $(x_i \geq 0, y_i \geq 0)$ and the same length $(i = 1, ..., n)$ it is evaluated as $J-T(\{x_i\}, \{y_i\}) = \sum_i \mbox{min}(x_i,y_i)/ \sum_i \mbox{max}(x_i,y_i)$. The maximum similarity is characterized by the J-T index equal to one, while very low similarity by the J-T index equal to zero. The very many applications of J-T index include, e.g. various forms of image analysis or identification of words that are mistyped by the users of numerous computer and cell-phone/smart-phone applications. 

The second measure is the scalar (inner) product of the vectors representing the set of measured probabilities. 
It is worth noting that while the vectors of measured probabilities have the sum of components equal to one, their length in a vector sense is in general not equal to one.
The length of such a vector is equal to one only when one of the basis states has the probability equal to one and all others have the probability equal to zero. 
It is also the maximum length in the vector sense. 
For any 
other distribution of probabilities the length is lower than one.
The other extreme case with the shortest length of the vector has all $n$ basis states with the same probability, equal to $1/n$ length, and the length of this vector is $1/\sqrt{n}$.
 Therefore, all the probability vectors are normalized  before we evaluate their scalar product in our analysis. 

\begin{figure}[ht]
	\centering
	\includegraphics[width=0.99\linewidth]{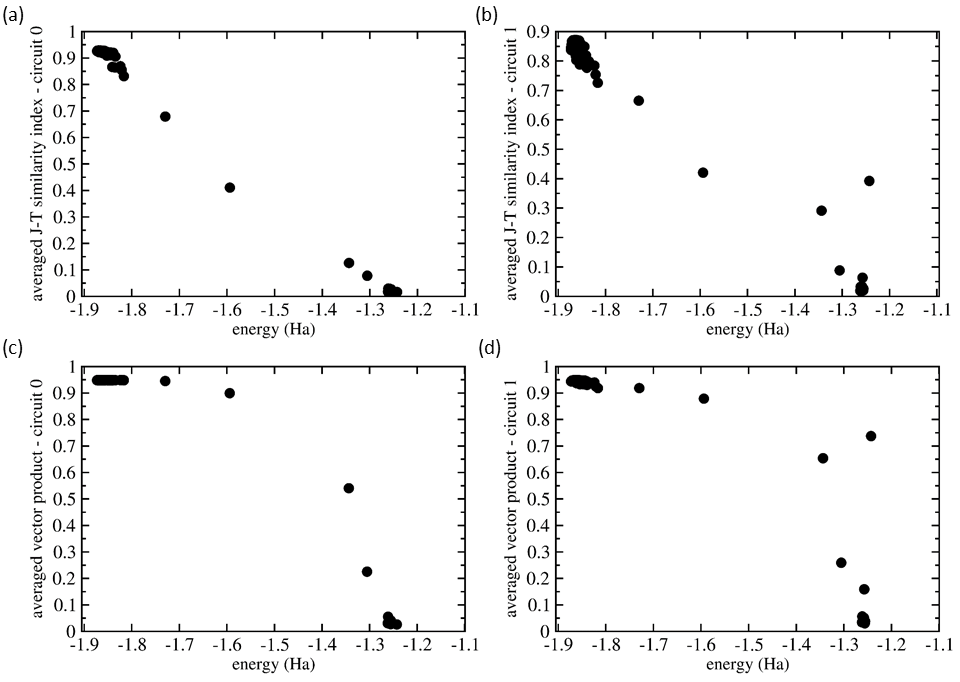}
	\caption{Comparison of similarities of vectors of measured probabilities as functions of energies. 
	The plotted values are averaged ones using all $200$ values and evaluated using either the Jaccard-Tanimoto (J-T) similarity index (parts (a,b)) and scalar product (parts (c,d)) for both the circuit $0$ (parts (a,c)) and circuit $1$ (parts (b,d)).}
	\label{sim-fig}
\end{figure}

As individual additive parts of the Hamiltonian can be divided into two groups with each requiring a different topology of the quantum circuit, we have evaluated similarities as obtained for two sets of $200$ vectors of measured probabilities that we below call circuit $0$ (all qubits measured in the computational basis) and circuit $1$ (qubits $1$ and $3$ measured in the computational basis and qubits $0$ and $2$ measured in the $X$ basis). 
Figure~\ref{sim-fig} shows the values of both the Jaccard-Tanimoto (J-T) similarity index (Fig.~\ref{sim-fig}(a,b)) and scalar product (Fig.~\ref{sim-fig}(c,d)) for both the circuit $0$ (Fig.~\ref{sim-fig}(a,c)) and circuit $1$ (Fig.~\ref{sim-fig}(b,d)). The values plotted for each vector 
in each set were determined as averaged values over $200$ J-T/scalar-product similarities of a given vector and all $200$ vectors in each set. 

Figure~\ref{sim-fig} shows that the averaged similarity values are decreasing with increasing energy 
from its ground-state value. 
Without noise, the probability vector of circuit $0$ has the $\ket{1100}$ basis state with probability close to $1$ and the other basis states have probability close to $0$. From the $200$ calculations, $188$ of them, i.e. $94\%$ , exhibit variations of this scenario and these results cover energies up to $-1.7$ Hartree. The J-T similarity index values (see Fig.~\ref{sim-fig}(a)) show a weakly decreasing trend with increasing energy. In contrast, the averaged scalar product of all of these $188$ probability vectors has nearly identical value (see Fig.~\ref{sim-fig}(c)). As another extreme, about $5\%$ of vectors of probabilities have very low averaged J-T similarity value close to zero and the energies from the range between $-1.3$ and $-1.2$ Hartree. These we, therefore, identify as excited states because their vectors of probabilities are very different from the case when the $|$1110$\rangle$ basis state has the probability  close to one. The remaining $1\%$ cases plotted in Fig.~\ref{sim-fig}(a,c) are characterized by energies and similarity index values that are in between the region close to the ground state on one hand and that of excited states on the other hand. 
We believe that these are essentially erroneous calculations in which the SPSA algorithm failed to converge to a local minimum.

The above mentioned percentages ($94\%$ states similar to the ground state, $5\%$ similar to the excited states and $1 \%$ erroneous states) explain the actual value of the averaged similarity index plotted in Figure~\ref{sim-fig}. As $94\%$ vectors of probability vectors are related to the ground state, they are mutually similar and contribute with values close to one. Their averaged similarity index is lower than one because it is reduced by their medium similarity to the  erroneous states ($1\%$ of contributions into the sum) and nearly zero similarity value with the excited states (another $5\%$ of contributions). Complementarily, the excited states have nearly zero similarity index with $94\%$ of all probability vectors as well as erroneous states and, therefore, their averaged similarity index is close to zero. Quite a similar situation is that of the intermediate values of the erroneous states that are similar to both the ground-state related states as well as the excited states. 

The situation is more complicated in the case of circuit $1$ (see Fig.~\ref{sim-fig}(b,d)) when the measurements of the ground state result in  outcomes ($|$0100$\rangle$, $|$0110$\rangle$, $|$1100$\rangle$ and $|$1110$\rangle$) with rather similar probabilities (they differ by about $1/16$). The  J-T similarity index performs quite similarly (see Fig.~\ref{sim-fig}(b)) as in the case of circuit $0$ (see Fig.~\ref{sim-fig}(a)) but some of the excited states show significant
non-zero values (up to $0.4$).
This trend is a lot more pronounced in the case of scalar product (see Fig.~\ref{sim-fig}(d)) when the values can reach as high as $0.74$.

While future studies focused on similarity analysis in other systems are critically needed, a few general conclusions can be drawn. 
First, the similarity analysis of vectors of measured probabilities can be very fruitful when assessing the computed energies. 
The similarities seem to be able to tell cases that are just noisy-variants of the ground state from those that are (i) related to other eigenvalues of higher-energy excited states or even (ii) plainly erroneous cases.
Second, the two tested similarity measures, the Jaccard-Tanimoto (J-T) similarity index and the scalar product, perform much better when the ground-state and/or excited states are associated with the situation when one of the basis states has probability close to $1$ while the others have it equal to $0$. 
It would be thus advantageous to develop methods associating the eigenvalues with these cases.
Third, while the similarity index based on the scalar product identifies states related to the ground state by noise a lot more clearly than
 the J-T index, it tends to overrate the similarity of the excited and plain erroneous states. 
Therefore, the J-T similarity index seems to better capture continuously varying levels of similarity of different states. 
Fourth, we have analyzed averaged similarity measures based on the similarity between each vector of probabilities with respect to all other such vectors in our data set. 
Our analysis does not profit from the fact that we can estimate the vector of probabilities related to the ground state (it has the lowest energy) and determine the similarity with respect to this particular state. 
As the VQE can be applied to determine also the excited states, the same procedure with determination of the similarity analysis can be applied also to the excited states and the their noise-related states can be more identified. 
Lastly, when an eigenvalue is associated with the probability vector where one of the basis states has the probability equal to one and others zero, all other vectors that are noise-related to it (and that have probabilities redistributed elsewhere due to the nose and errors),  can be used for determination for effective error rates and reversely applied for correcting for these errors.

\section{Data tables}
\label{AppendixDataTables}
Let us now present the data behind the figures in the main text. First, in Table~\ref{tab:maxiter1000}, we list the median ground state energies and the percentages of data falling into chemical precision range (with and without precise recalculation of energies), for various settings of the number of \texttt{shots} and \texttt{maxiter} = $1000$.
This is followed by Tables~\ref{tab:QuantumUses} and \ref{tab:QuantumUses2} in which we elaborete in more detail on energy values obtainable by various \texttt{shots} and \texttt{maxiter} settings. Then, in Table~\ref{tab3}, we list the results underlying Figure~\ref{fig:forms}, comparing state preparation ansatzes for $2$- and $4$-qubit computations.
Further, in Table~\ref{tab2}, we list the results highlighting the variability of actual hardware when performing runs on different dates, plotted in Figure~\ref{fig:errors}.
Finally, in Table~\ref{tab:realRun} we show energies underlying real runs presented in Figure~\ref{fig:bogota}.

\begin{specialtable}[ht]
\widefigure
\centering
    \begin{tabular}{ccccccc}
    \toprule
\multirow{2}{*}{Simulator}                        & \multirow{2}{*}{\texttt{shots}}    & \multirow{2}{*}{Median [Ha]}   & \multirow{2}{*}{$\%$}     &  \multicolumn{2}{c}{Recalculation} \\
&&&& Median [Ha]   & $\%$ \\ \midrule
\multirow{5}{*}{qasm\_simulator} & $512$             & $ -1.86637 $ & $ 10.5 \pm 3 $  & $ -1.86675 $ & $ 93.8 \pm 3 $\\
                                 & $1024$            & $ -1.86723 $ & $ 14.8 \pm 3 $  & $ -1.86693 $ & $ 98.5 \pm 3 $\\
                                 & $2048$            & $ -1.86667 $ & $ 21.1 \pm 3 $  & $ -1.86703 $ & $ 99.1 \pm 3 $\\
                                 & $4096$            & $ -1.86705 $ & $ 28.7 \pm 3 $  & $ -1.86707 $ & $ 99.0 \pm 3 $\\
                                 & $8192$            & $ -1.86729 $ & $ 38.9 \pm 3 $  & $ -1.86710 $ & $ 99.4 \pm 3 $\\
statevector\_simulator           & $-$               & $ -1.86712 $ & $ 99.1 \pm 3 $  & $-$ & $-$\\ \bottomrule
\end{tabular}
\caption{Summary of the medians of $1000$ optimized ground state energies and the percentage of outcomes that were found within chemical precision  of exact energy ($-1.86712 \pm 0.0015$ Ha) for different settings of \texttt{shots} parameter and \texttt{maxiter} = $1000$. Additionally, in the middle two columns the final energy is the direct outcome of the SPSA algorithm and in the last two columns recalculated using statevector simulator.
The $R_{\mathrm{y}}$ variational ansatz was used. Error margin of $3$ percent is estimated based on the standard deviation of $1000$ identically distributed trials. This data is the basis for Figure \ref{fig:shots}.}
\label{tab:maxiter1000}
\end{specialtable}

\begin{specialtable}[ht]
\widefigure
\centering
    \begin{tabular}{cccccc}
    \toprule
Settings                             & \texttt{shots} & Median [Ha]   & $\%$   \\ \midrule
\multirow{4}{*}{SPSA(\texttt{maxiter} = 50)} & $512$         &    $ -1.86119 $ & $ 8.6 \pm 3 $  \\
                                     & $1024$            &  $ -1.86258 $ & $ 11.4 \pm 3 $ \\
                                     & $4096$            &  $ -1.86475 $ & $ 20.7 \pm 3 $ \\
                                     & $8192$            &  $ -1.86544 $ & $ 29.9 \pm 3 $ \\ \midrule
\multirow{4}{*}{SPSA(\texttt{maxiter} = 75)} & $512$         &    $ -1.86276 $ & $ 9.1 \pm 3 $  \\
                                     & $1024$            &  $ -1.86439 $ & $ 12.6 \pm 3 $ \\
                                     & $4096$            &  $ -1.86559 $ & $ 24.9 \pm 3 $ \\
                                     & $8192$            &  $ -1.86601 $ & $ 34.2 \pm 3 $ \\ \midrule
\multirow{4}{*}{SPSA(\texttt{maxiter} = 100)} & $512$        &     $ -1.86419 $ & $ 9.4 \pm 3 $  \\
                                     & $1024$            &  $ -1.86530 $ & $ 13.5 \pm 3 $ \\
                                     & $4096$            &  $ -1.86607 $ & $ 25.9 \pm 3 $ \\
                                     & $8192$            &  $ -1.86644 $ & $ 38.1 \pm 3 $ \\ \midrule
\multirow{4}{*}{SPSA(\texttt{maxiter} = 125)} & $512$        &     $ -1.86486 $ & $ 9.3 \pm 3 $  \\
                                     & $1024$            &  $ -1.86524 $ & $ 13.6 \pm 3 $ \\
                                     & $4096$            &  $ -1.86631 $ & $ 27.0 \pm 3 $ \\
                                     & $8192$            &  $ -1.86656 $ & $ 37.5 \pm 3 $ \\ \midrule
\multirow{4}{*}{SPSA(\texttt{maxiter} = 150)} & $512$        &     $ -1.86461 $ & $ 9.3 \pm 3 $  \\
                                     & $1024$            &  $ -1.86593 $ & $ 14.3 \pm 3 $ \\
                                     & $4096$            &  $ -1.86613 $ & $ 28.8 \pm 3 $ \\
                                     & $8192$            &  $ -1.86637 $ & $ 36.9 \pm 3 $ \\ \midrule
\multirow{4}{*}{SPSA(\texttt{maxiter} = 200)} & $512$        &     $ -1.86634 $ & $ 11.3 \pm 3 $  \\
                                     & $1024$            &  $ -1.86575 $ & $ 14.3 \pm 3 $ \\
                                     & $4096$            &  $ -1.86652 $ & $ 28.7 \pm 3 $ \\
                                     & $8192$            &  $ -1.86653 $ & $ 36.5 \pm 3 $ \\ \bottomrule
\end{tabular}
\caption{Summary of the medians of $1000$ optimized ground state energies and the percentage of outcomes that were found within chemical precision  of exact energy ($-1.86712 \pm 0.0015$ Ha) for different settings of maximum number of iterations (\texttt{maxiter}) and number of shots. The $R_{\mathrm{y}}$ variational ansatz was used. Error margin of $3$ percent is estimated based on the standard deviation of $1000$ identically distributed trials. This data is the basis for Figure \ref{fig:maxiter}.
\label{tab:QuantumUses}
}
\end{specialtable}

\begin{specialtable}[ht]
\widefigure
\centering
    \begin{tabular}{ccccccc}
    \toprule
\multirow{2}{*}{Settings}                        & \multirow{2}{*}{\texttt{shots}}    & \multirow{2}{*}{Median [Ha]}   & \multirow{2}{*}{$\%$}     &  \multicolumn{2}{c}{Recalculation} \\
&&&& Median [Ha]   & $\%$ \\ \midrule
\multirow{2}{*}{SPSA(\texttt{maxiter} = 50)} & $512$ & $ -1.86101 $ & $ 7.0 \pm 3 $  & $ -1.86349 $ & $ 25.5 \pm 3 $\\
                                 & $1024$            & $ -1.86190 $ & $ 10.8 \pm 3 $ & $ -1.86520 $ & $ 41.4 \pm 3 $\\ \midrule
\multirow{2}{*}{SPSA(\texttt{maxiter} = 75)} & $512$ & $ -1.86369 $ & $ 9.7 \pm 3 $ & $ -1.86511 $ & $ 38.7 \pm 3 $\\
                                 & $1024$            & $ -1.86420 $ & $ 11.7 \pm 3 $ & $ -1.86594 $ & $ 57.4 \pm 3 $\\ \midrule
\multirow{2}{*}{SPSA(\texttt{maxiter} = 100)}& $512$ & $ -1.86472 $ & $ 9.7 \pm 3 $ & $ -1.86560 $ & $ 49.4 \pm 3 $\\
                                & $1024$            & $ -1.86458 $ & $ 13.9 \pm 3 $ & $ -1.86627 $ & $ 64.5 \pm 3 $\\ \bottomrule
\end{tabular}
\caption{Summary of the medians of $1000$ optimized ground state energies and the percentage of outcomes that were found within chemical precision  of exact energy ($-1.86712 \pm 0.0015$ Ha) for different settings of maximum number of iterations (\texttt{maxiter}) and number of shots. The $R_{\mathrm{y}}$ variational ansatz was used. Error margin of $3$ percent is estimated based on the standard deviation of $1000$ identically distributed trials. Rightmost two collumns are the recalculation of obrtained energies using statevector simulator.
This data is the basis for Figure \ref{fig:maxiter2}.\label{tab:QuantumUses2}}
\end{specialtable}



\begin{specialtable}[ht]
\widefigure
\centering
\begin{tabular}{cccc}
\toprule
    \multicolumn{4}{c}{2-qubit   Hamiltonian H$_{\mathrm{2}}$}                    \\
    Form & Depth (parameters) & Median [Ha]   & $\%$     \\ 
    \midrule
    $R_\mathrm{y}$               & $3$ $(8)$              &    $-1.86713$ & $31.2\pm 3$     \\
    $R_\mathrm{y}$$R_\mathrm{z}$             & $2$ $(12)$             & $-1.86664$ & $30.7\pm 3$ \\
    $R_\mathrm{y}$               & $2$ $(6)$              & $-1.86709$ & $32.4\pm 3$   \\
    $R_\mathrm{y}$$R_\mathrm{z}$             & $1$ $(8)$              & $-1.86594$ & $25.8\pm 3$ \\
    $R_\mathrm{y}$               & $1$ $(4)$              & $-1.86705$ & $28.7\pm 3$    \\ 
    \toprule
    \multicolumn{4}{c}{4-qubit Hamiltonian H$_{\mathrm{2}}$}             \\
    Form & Depth (parameters) & Median [Ha]   & $\%$   \\ 
    \midrule
    $R_\mathrm{y}$               & $5$ $(24)$             & $-1.86331$ & $18.1 \pm 3$ \\
    $R_\mathrm{y}$$R_\mathrm{z}$             & $2$ $(24)$             & $-1.86070$ & $12.5\pm 3$ \\
    $R_\mathrm{y}$               & $2$ $(12)$             & $-1.86361$ & $17.9\pm 3$ \\
    $R_\mathrm{y}$$R_\mathrm{z}$             & $1$ $(16)$             &  $-1.84560$ & $0.0$    \\
    $R_\mathrm{y}$               & $1$ $(8)$              & $-1.84591$ & $0.0$ \\ 
    \bottomrule
    \end{tabular}
\caption{A comparison of various settings of quantum circuits for 2-qubit and 4-qubit Hamiltonians of H$_{\mathrm{2}}$. The $R_{\mathrm{y}}R_{\mathrm{z}}$ and $R_{\mathrm{y}}$ variational forms accompanied by linear entanglement with different depth of circuit and number of parameters were used. In case of 2-qubit Hamiltonian we used unrestricted maximum number of iterations of the SPSA optimizer, and in case of 4-qubit Hamiltonian the \texttt{maxiter} was set to 400.
Columns labeled $\%$ represent the percentage of the runs that ended within chemical accuracy ($-1.86712 \pm 0.0015$ Ha) of the optimum. Error margin of $3$ percent is estimated based on the standard deviation of $1000$ identically distributed trials. This data is the basis for Figure \ref{fig:forms}.
\label{tab3}}
\end{specialtable}

\begin{specialtable}[ht]
\widefigure
\centering
\begin{tabular}{ccccccc}
\toprule 
\multirow{2}{*}{\begin{tabular}[c]{@{}c@{}}Simulator \\ NoiseModel\end{tabular}} & \multirow{2}{*}{Qubits} & \multirow{2}{*}{Date} & \multicolumn{2}{c}{$R_\mathrm{y}$ form} & \multicolumn{2}{c}{$R_\mathrm{y}$$R_\mathrm{z}$  form} \\
                           &                                   &                       & Median[Ha]               & \%               & Median[Ha]                  & \%                \\ \midrule
\multirow{3}{*}{
gate errors}      & \multirow{2}{*}{2}                & Dec.\,14,20      & $-1.85819$ & $3.3\pm 3$              & $-1.85719$ & $2.4\pm 3$               \\
                           &                                   & May\,14,21           &                        &                  & $-1.85947$ & $5.3\pm 3$              \\
                           & $4$                                 & Dec.\,14,20      & $-1.79786$ & $0.0$               & $-1.77738$                & $0.0$                 \\ \midrule
\multirow{3}{*}{\begin{tabular}[c]{@{}c@{}}readout\\ errors\end{tabular}}   & \multirow{2}{*}{2}                & Dec.\,14,20      & $-1.80617$ & $0.0$               & $-1.80510$ & 0.0               \\
                           &                                   & May\,14,21           &                        &                  & $-1.82060$ & $0.0$                \\
                           & $4$                                 & Dec.\,14,20      & $-1.78270$ & $0.0$                & $-1.76214$                & $0.0$                 \\ \midrule
\multirow{3}{*}{
all errors}       & \multirow{2}{*}{2}                & Dec.\,14,20   &   $-1.79816$ & $0.0$               & $-1.79646$ & $0.0$                 \\
                           &                                   & May\,14,21           &                        &                  & $-1.81879$ & $0.0$                 \\
                           & $4$                                 & Dec.\,14,20      & $-1.70967$         & $0.0$                & $-1.68760 $               & $0.0$              \\ \bottomrule
\end{tabular}
\caption{A comparison of results from different dates and different noise models built from a real quantum processor ibmq$\_$santiago. The number of shots was set to $4096$. The calculations were performed using both the $R_{\mathrm{y}}R_{\mathrm{z}}$ and $R_{\mathrm{y}}$ variational forms. The classical optimization method SPSA was used, for $2$-qubit system the maxiter was unrestricted and for $4$-qubit system, the maxiter was set to $400$. Last column represents the percentage of the runs that ended within chemical accuracy ($-1.86712 \pm 0.0015$ Ha) of the optimum. Error margin of $3$ percent is estimated based on the standard deviation of $1000$ identically distributed trials. This data is the basis for  Figures \ref{fig:12} and \ref{fig:errors}.\label{tab2}}
\end{specialtable}

\begin{specialtable}[ht]
\widefigure
\centering
    \begin{tabular}{ccc}
    \toprule
 \multicolumn{2}{c}{\em ibm\_lagos}                      &  \multirow{2}{*}{\begin{tabular}[c]{@{}c@{}}Statevector \\ simulator\end{tabular}}   \\
\texttt{shots} = $1024$   & \texttt{shots} = $40000$\\   \midrule
$-1.85075$ Ha & $-1.85688$ Ha &  $-1.86588$ Ha \\
$-1.86237$ Ha & $-1.86035$ Ha &  $-1.86558$ Ha \\
$-1.81998$ Ha & $-1.82370$ Ha &  $-1.86560$ Ha \\
$-1.82088$ Ha & $-1.82312$ Ha &  $-1.85678$ Ha \\
$-1.84165$ Ha & $-1.83561$ Ha &  $-1.86488$ Ha \\
$-1.82453$ Ha & $-1.83271$ Ha &  $-1.86456$ Ha \\
$-1.85601$ Ha & $-1.86479$ Ha &  $-1.86569$ Ha \\
$-1.87990$ Ha & $-1.86107$ Ha &  $-1.86605$ Ha \\
$-1.84045$ Ha & $-1.81550$ Ha &  $-1.86504$ Ha \\
$-1.86599$ Ha & $-1.82214$ Ha &  $-1.86554$ Ha \\
$-1.78271$ Ha & $-1.84946$ Ha &  $-1.86099$ Ha \\
$-1.82068$ Ha & $-1.83412$ Ha &  $-1.86482$ Ha \\
$-1.83439$ Ha & $-1.82281$ Ha &  $-1.85480$ Ha \\
$-1.80671$ Ha & $-1.83176$ Ha &  $-1.86270$ Ha \\ \bottomrule
\end{tabular}
\caption{Energy values obtained by $ibm\_lagos$ with various final energy evaluations. This data is the basis for Figure \ref{fig:bogota}.\label{tab:realRun}}
\end{specialtable}



\end{paracol}
\end{document}